\documentclass[12pt,preprint]{aastex} 

\newcommand{\etal}{{\it{et al.}}~}
\newcommand{\ie}{{\it{i.e.}}~}
\newcommand{\eg}{{\it{e.g.}}}

\newcommand{\chisq}{${\chi}^2$~}

\begin{document}

\title{The Sloan Digital Sky Survey-II: Photometry and Supernova Ia Light Curves from the 2005 data}

\shorttitle{SDSS-II Supernova Photometry}
\shortauthors{Holtzman \etal}

\author{
Jon~A.~Holtzman,\altaffilmark{1}
John~Marriner,\altaffilmark{2}
Richard~Kessler,\altaffilmark{3,4}
Masao~Sako,\altaffilmark{5,6}
Ben~Dilday,\altaffilmark{3,7}
Joshua~A.~Frieman,\altaffilmark{2,3,8}
Donald~P.~Schneider,\altaffilmark{9}
Bruce~Bassett,\altaffilmark{10,11}
Andrew~Becker,\altaffilmark{12}
David~Cinabro,\altaffilmark{13}
Fritz~DeJongh,\altaffilmark{3}
Darren~L.~Depoy,\altaffilmark{14}
Mamoru~Doi,\altaffilmark{15}
Peter~M.~Garnavich,\altaffilmark{16}
Craig~J.~Hogan,\altaffilmark{12}
Saurabh~Jha,\altaffilmark{5,17}
Kohki Konishi,\altaffilmark{18}
Hubert~Lampeitl,\altaffilmark{19,20}
Jennifer~L.~Marshall,\altaffilmark{14}
David~McGinnis,\altaffilmark{3}
Gajus~Miknaitis,\altaffilmark{3}
Robert~C.~Nichol,\altaffilmark{20}
Jose~Luis~Prieto,\altaffilmark{14}
Adam~G.~Reiss,\altaffilmark{19,21}
Michael~W.~Richmond,\altaffilmark{22}
Roger~Romani,\altaffilmark{5}
Mathew~Smith,\altaffilmark{20}
Naohiro~Takanashi,\altaffilmark{15}
Kouichi~Tokita,\altaffilmark{15}
Kurt~van~der~Heyden,\altaffilmark{11,23}
Naoki~Yasuda,\altaffilmark{18}
Chen~Zheng\altaffilmark{5}
}

\altaffiltext{1}{
  Department of Astronomy,
   MSC 4500,
   New Mexico State University, P.O. Box 30001, Las Cruces, NM 88003,
   holtz@nmsu.edu
}
\altaffiltext{2}{
Center for Particle Astrophysics, 
  Fermi National Accelerator Laboratory, P.O. Box 500, Batavia, IL 60510.
}
\altaffiltext{3}{
  Kavli Institute for Cosmological Physics, 
   The University of Chicago, 5640 South Ellis Avenue Chicago, IL 60637.
}
\altaffiltext{4}{
Enrico Fermi Institute,
University of Chicago, 5640 South Ellis Avenue, Chicago, IL 60637.
}
\altaffiltext{5}{
 Kavli Institute for Particle Astrophysics \& Cosmology, 
  Stanford University, Stanford, CA 94305-4060.
}
\altaffiltext{6}{
Department of Physics and Astronomy,
University of Pennsylvania, 203 South 33rd Street, Philadelphia, PA  19104.
}
\altaffiltext{7}{
Department of Physics, 
University of Chicago, Chicago, IL 60637.
}
\altaffiltext{8}{
  Department of Astronomy and Astrophysics,
   The University of Chicago, 5640 South Ellis Avenue, Chicago, IL 60637.
}
\altaffiltext{9}{
  Department of Astronomy and Astrophysics,
   The Pennsylvania State University,
   525 Davey Laboratory, University Park, PA 16802.
}
\altaffiltext{10}{
Department of Mathematics and Applied Mathematics,
University of Cape Town, Rondebosch 7701, South Africa.
}
\altaffiltext{11}{
  South African Astronomical Observatory,
   P.O. Box 9, Observatory 7935, South Africa.
}
\altaffiltext{12}{
  Department of Astronomy,
   University of Washington, Box 351580, Seattle, WA 98195.
}
\altaffiltext{13}{
Department of Physics, 
Wayne State University, Detroit, MI 48202.
}
\altaffiltext{14}{
  Department of Astronomy,
   Ohio State University, 140 West 18th Avenue, Columbus, OH 43210-1173.
}
\altaffiltext{15}{
  Institute of Astronomy, Graduate School of Science,
   University of Tokyo 2-21-1, Osawa, Mitaka, Tokyo 181-0015, Japan.
}
\altaffiltext{16}{
  University of Notre Dame, 225 Nieuwland Science, Notre Dame, IN 46556-5670.
}

\altaffiltext{17}{
Department of Physics and Astronomy, 
Rutgers University, Piscataway, NJ08854.
}
\altaffiltext{18}{
Institute for Cosmic Ray Research,
University of Tokyo, 5-1-5, Kashiwanoha, Kashiwa, Chiba, 277-8582, Japan.
}
\altaffiltext{19}{
  Space Telescope Science Institute,
   3700 San Martin Drive, Baltimore, MD 21218.
}
\altaffiltext{20}{
  Institute of Cosmology and Gravitation,
   Mercantile House,
   Hampshire Terrace, University of Portsmouth, Portsmouth PO1 2EG, UK.
}
\altaffiltext{21}{
Department of Physics and Astronomy,
Johns Hopkins University, 3400 North Charles Street, Baltimore, MD 21218.
}
\altaffiltext{22}{
  Physics Department,
   Rochester Institute of Technology,
   85 Lomb Memorial Drive, Rochester, NY 14623-5603.
}
\altaffiltext{23}{
Department of Astronomy,
University of Cape Town, South Africa.
}

\altaffiltext{24}{
  Department of Astronomy,
   McDonald Observatory, University of Texas, Austin, TX 78712
}
\altaffiltext{25}{
  Apache Point Observatory, P.O. Box 59, Sunspot, NM 88349.
}
\altaffiltext{26}{
Department of Astronomy,
Seoul National University, Seoul, South Korea.
}
\altaffiltext{27}{
Department of Astronomy,
Columbia University, New York, NY 10027.
}
\altaffiltext{28}{
Lowell Observatory, 1400 Mars Hill Rd., Flagstaff, AZ 86001
}
\altaffiltext{29}{
Subaru Telescope, 650 N. A'Ohoku Place, Hilo, HI 96720
}
\altaffiltext{30}{
  Obserwatorium Astronomiczne na Suhorze,
   Akademia Pedagogicazna w Krakowie,
   ulica Podchor\c{a}\.{z}ych 2, PL-30-084 Krak\'{o}w, Poland.
}
\altaffiltext{31}{
Gemini Observatory, 670 North A'ohuoku Place, Hilo, HI 96720.
}

\begin{abstract}
We present \textit{ugriz} light curves for 146 spectroscopically confirmed or 
spectroscopically probable Type Ia supernovae from the 2005 season of the
SDSS-II Supernova survey.  The light curves have been constructed using
a photometric technique that we call scene modelling, which is described
in detail here; the major feature is that supernova brightnesses are
extracted from a stack of images without spatial resampling or convolution
of the image data.  This procedure produces accurate photometry along
with accurate estimates of the statistical uncertainty, and can be used
to derive photometry taken with multiple telescopes. We discuss
various tests of this technique that demonstrate its capabilities. We
also describe the methodology used for the calibration of the photometry,
and present calibrated magnitudes and fluxes for all of the spectroscopic 
SNe Ia from the 2005 season.

\end{abstract}


\section{Introduction}

In its second phase of operations, the Sloan Digital Sky Survey (SDSS;
York \etal 2000) telescope has been used to attack several different
scientific programs.  One of these is a supernova survey that ran
from 1 September to 30 November for three years (2005-2007) and 
targetted Type Ia supernovae in the redshift range $0.05<z<0.35$. The project's
scientific motivations are 1) to take advantage of the high areal
coverage (over 150 square degrees per night) and moderate sensitivity
($\sim$ 22 mag) that can be obtained with the large format camera and drift
scanning of SDSS to fill in a redshift regime where other surveys have
not been efficient in finding supernovae, and 2) to take advantage of
the well-understood photometric system of SDSS to minimize calibration
errors and other systematics. An overview of the observational techniques
and expected scientific returns of this program are given in 
Frieman \etal (2008).

Operationally, two strips (denoted strips 82N and 82S) located along
the celestial equator with right ascension between 20h and 4h are
monitored over a period of three months from September through November.
These two strips, with a combined width of $2.5^{\circ}$ and an area of
approximately 300 square degrees, have been the subject of many previous
SDSS imaging scans during the original SDSS survey. The SDSS SN survey
alternates between these two strips on successive clear nights. There is
a small overlap between the strips (roughly 10\% of the area) to insure no
sky coverage is lost.  New transients and variable sources are identified
by subtracting high signal-to-noise (S/N) template images constructed by
coadding previous observations of the strip and inspecting the subtracted
frames to find new objects. Candidate supernovae are identified via
a combination of automated and interactive techniques, and observed
spectroscopically using a variety of telescopes to confirm that they are
supernova and to determine the supernova type and redshift.  Details of
the supernovae candidate identification and spectroscopic target selection
are given in Sako \etal (2008); details of the spectroscopy and supernova
typing are given in Zheng \etal (2008). The initial cosmological results
from the SDSS-II Supernova Survey are presented in Kessler \etal (2009).

This paper presents the techniques used to measure the brightnesses of
the supernovae for final analysis. We discuss the photometric calibration,
photometric techniques, and expected errors in the resulting photometry,
and present the resulting light curves for spectroscopically confirmed
and likely type Ia supernovae from the first season (2005) of the SDSS-II
Supernova Survey.

\section{Data and data reduction}

\subsection{SDSS imaging}

The imaging data are taken using the SDSS imaging camera (Gunn \etal
1998) on the SDSS 2.5m telescope (Gunn \etal 2006) at Apache Point
Observatory (APO).  This camera uses 30 imaging CCDs arranged in
6 columns; each column has a detector for each of the 5 SDSS filter
bandpasses, \textit{ugriz} (Fukugita \etal 1996). Additional detectors 
are used to assist with
the astrometric calibration (Pier \etal 2003) of the science frames.
The camera runs in drift scanning mode such that each column is exposed
for 54 seconds per filter, with a slight time lag between successive
filters.  Operation of the camera for the SDSS-II supernova survey is
identical to routine operation for the original survey (see Stoughton
\etal 2002 and Adelman-McCarthy \etal 2008 for details of the survey
operation and data releases).

The imaging data are processed through the standard SDSS processing
pipeline, which uses the program PHOTO (Lupton \etal 2008) to remove
instrumental signatures, flag bad pixels, determine a PSF, and create
an object catalog with instrumental brightnesses. As output, PHOTO
produces corrected frames, which have instrumental signatures removed
and astrometric information in their headers, and mask frames that flag
problematic pixels. Each column of a strip in the sky is divided into
a series of adjacent fields (2048 $\times$ 1489 pixels, or roughly 800
by 600 arcsec), for output, with a small amount of overlap between fields.

For the purpose of identifying supernovae during the survey, template
images from previous imaging scans are subtracted from the images from
each supernova run. For the 2005 observing season, we used 
data from pre-2004 SDSS runs to create a coadded template. These
coadded templates were constructed from a combination of between 4 and
9 photometric runs with good seeing. Before subtracting the template,
a smearing kernel is applied to the template to match its PSF to the
PSF of the supernova frame, and the template frame is astrometrically
registered to the supernova frame.  Details of the astrometric and PSF
matching are given in an appendix in Sako \etal (2008). We refer to the
resulting subtracted frames as the {\tt Framesub} data. These data are used
for identification of candidate supernovae, and for initial photometry
that is used for target selection for spectroscopic followup; our
final photometry, as discussed below, is more accurate, but is not
used for target selection.

\subsection{Other imaging}

Additional imaging of SDSS-II SN candiates was obtained from several other
telescopes: primarily, the 2.4m MDM telescope on Kitt Peak, the 88in UH
telescope on Mauna Kea, and the ARC 3.5m and NMSU 1m telescopes at Apache
Point Observatory.  The main goals of these observations were to increase
light curve coverage during periods of poor weather that limited the
temporal coverage of the SDSS 2.5m data, to allow deeper observations
for more distant SN and/or at later epochs than can be obtained with
the fixed 54s integration time of the 2.5m telescope, and to measure the light
curves of supernovae discovered late in the survey season into the month
following the completion of the SDSS imaging. During the 2005 campaign,
the APO weather was generally quite good, so these additional observations
were not as critical as they might have been in poor weather.

On all of the non-SDSS telescopes, filter sets approximating the
SDSS filter set were used, but there are still small but significant
differences between the response functions.  This is a serious issue
for the supernova program, since we hope to reduce the systematic errors in the
photometry to $\sim$ 0.01 magnitudes.  In section \ref{sect:non25m},
we discuss the techniques used to extract the supernova photometry from
these other telescopes and the issues involved with using this photometry
in a joint analysis with the 2.5m data.

\section{Photometric calibration}

The SDSS-II supernovae runs are taken on all fall nights during which the
telescope can be operated, except for five nights around full moon. Much
of the data is taken under non-photometric conditions. However, all of
the data on strips 82N and 82S taken as part of the standard SDSS-I survey
(before 2004) were taken under photometric conditions, with simultaneous
monitoring of atmospheric transmission using the SDSS Photometric 
Telescope (PT).  As a result, the standard SDSS-I imaging provides multiple
photometric measurements of all stars along these strips.  The details
of the photometric calibration of the SDSS images is discussed in Hogg
\etal (2001), Smith \etal (2002), Ivezic \etal (2004), and Tucker \etal
(2006), and on the SDSS Web pages (http://www.sdss.org).

Ivezic \etal (2007) have taken the repeat observations along the
equatorial strips and constructed a master catalog of standard stars in
the SDSS system using these measurements.  Variable stars are flagged
by comparing the multiple measurements, and final median magnitudes
for all non-variables with good S/N were compiled into the master
catalog. A variety of tests with these measurements suggest that the
catalog magnitudes are accurate to within $\sim$ 0.01 magnitudes.

We use the Ivezic \etal (2007) catalog to calibrate the supernovae
frames. The details of which stars are used varies for the different
photometric techniques, as discussed below, but in general, brightness
measurements of a set of stars are made around each supernova candidate,
and these measurements are compared with the catalog to determine
photometric zeropoints for measurements of that candidate. Along with
the derived zeropoints, the scatter of the observed star brightnesses
relative to the catalog brightnesses are computed to determine how well
single zeropoints match the frames; with the drift scanning that is
used for the survey, stars at different right ascensions are observed
over different time intervals, so the zeropoint can vary as a function
of row position on the frames.

The number of calibration stars varies along the SDSS SN strip, largely
due to the variation in Galactic latitude. The number of calibration stars
around each supernova varies from a few to several hundred. However, in many cases,
a large fraction of the calibration stars do not have $u$-band magnitudes
in the calibration catalog, which limits our ability to extract $u$-band
measurements for some objects.

Finally, the Ivezic \etal (2007) catalog does not quite extend to the
western end of the supernova strip, in the first 10 degrees of the supernova
strip, because only a smaller number of SDSS runs covered this area. In this 
region, we have constructed an anagolous calibration catalog, but since
it is based on fewer observations, the uncertainties in the calibration
are a bit higher in this region.

\subsection{Absolute flux calibration}

\label{sect:fluxcal}

The Ivezic \etal (2007) catalog is calibrated to the native SDSS 
survey photometric system. While this system was originally 
intended to be an AB system (Oke 1974, Fukugita \etal 1996), it was
realized that the inclusion of the effects of atmospheric transmission
make it differ slightly (at about a 4 percent level) from an AB system
in the $u$-band; subsequent observations of calibrated targets suggest
that, at the 1-2 percent level, the survey photometry may differ from
that of a true AB system in the other bandpasses as well.

Various efforts have been made to understand the absolute calibration
of the SDSS system. Here, we calibrate to the HST white dwarf system
(Bohlin 2006). Bohlin (2001, 2004a, 2004b) present calibrated spectra for
several white dwarfs and solar analog stars on this system. Unfortunately,
all of these stars are too bright to be directly observed using the SDSS
2.5m telescope, and, in any case, none of them are in the region of the
sky where multiple SDSS observations have been made. However, several
of these stars have been observed numerous times by the SDSS Photometric
Telescope (PT), which is normally used to transfer photometric zeropoints
to data taken with the SDSS 2.5m telescope. While SDSS-like filters are
used on the PT, the system response functions are not exactly the same
between the two telescopes, so color terms have been determined to allow
for the transformation of magnitudes observed on the PT to the SDSS system
(which is defined as the system of the 2.5m telescope).  These color terms
have been defined over a relatively narrow range of color, corresponding
to F and G type stars. As a result, while the color terms do not strictly
apply to the white dwarf standards, the solar analog standards fall nicely
within the color range for which the color terms have been determined.

There are three solar analogs for which 10 or more PT observations have
been made: P330E, P177D, and P041C. The observed PT measurements were
transformed to the SDSS system using the standard survey color terms
(Tucker \etal, 2006). These SDSS measurements were then compared with
synthetic AB magnitudes calculated using the calibrated spectral energy
distributions (SEDs) of the stars and the SDSS system response curves
(from the SDSS web site, www.sdss.org).  Differences between
the synthetic and observed magnitudes are then interpreted to be the
deviation of the SDSS system from a true AB system. The average magnitude
offsets (AB-SDSS) for the three stars are determined to be $-0.037$, $0.024$,
$0.005$, $0.018$, and $0.016$ mag for \textit{ugriz}, respectively.  
Table \ref{tab:ab}
summarizes the observed and synthetic magnitudes for the solar analogs,
and the average offsets; the offsets are defined such that they need to
be added to the SDSS magnitudes to bring them onto an AB system.

We adopt these offsets for our supernova photometry, since accurate
absolute calibration is important for cosmological analysis of the
supernovae data. Note that these offsets rest on the assumptions of:
1) correct SEDs for the solar analogs, 2) correct observations of the
solar analogs, 3) correct transformations of the observations to the
SDSS system, and 4) correct knowledge of the SDSS system response.

We recognize that further refinements to the absolute calibration may
be available in the future. We note that several other efforts have
been made to understand the relation of the SDSS system to an AB
system (see SDSS web site) that yield results similar, but not identical,
to those adopted here.  Differences in these analyses at the 1-2 percent
level are consistent with our calibration error estimate of about 1\%.

Because of potential refinements to the absolute flux calibration,
we present two versions of supernova photometry for the data associated
with this paper: magnitudes on the native SDSS system (no AB correction), 
and fluxes that have been determined using the AB corrections discussed 
above.

%

\section{Photometry methods}

After the images are taken, initial, rapid photometry is required to
identify candidates for spectroscopic followup (Sako \etal 2008).  This quick
photometry, which we call search photometry, measures SN brightnesses
using a modified DOPHOT (Schechter, Mateo, \& Saha 1993) technique on
the pipeline template-subtracted frames.  Each observation in each filter
is processed independently.  Objects are not required to be present at a
common position in all filters and epochs and may be found in some filters
but not others.  The initial search photometry meets the goal of
supernova detection and measurement (generally much better than the 10\%
accuracy goal), but it does not provide the most accurate treatment of
the data possible.

For the final photometry, we investigated three different techniques. The
first, which we call forced photometry, also works on the template
subtracted frames, but the photometry reduction forces the position of
the supernova to be the same on all frames,  where the forced position
is determined from the average of the search photometry positions in
frames where the SN is within 1 magnitude of its peak observed brightness.
Forced photometry is used during the supernova search to obtain photometry
on supernova candidates for epochs and filters in which an object was
not detected by the initial photometry.  For both search and forced
photometry, the astrometric and photometric scalings of each frame are
adopted from the {\tt Framesub} software.

Two independent techniques that recompute the astrometric and photometric
scalings, as well as provide independent photometry on the supernovae, were
also developed. One, which we call ``cross-convolution'' photometry, 
measures stellar positions and intensities on search and template frames,
and determines an astrometric solution and a photometric scaling.
The template frame is convolved with the PSF of the search frame, and
the search frame is convolved with the PSF of the template frame; this
avoids the requirement of parameterizing the PSF as is done in frame
subtraction pipeline. The convolved template frame is subtracted from
the convolved search frame, and the magnitudes are determined by
weighted PSF photometry on the difference, again requiring the same 
position for the supernova in all frames. The cross-convolution photometry
uses PSFs as measured by the PHOTO pipeline.

Finally, we developed a technique that does not use template-subtracted
frames, but instead fits all of the individual reduced frames with a
model of the galaxy background and supernova; we call this technique
``scene-modelling'' photometry.   Ultimately, we chose to use ``scene
modeling" as the final photometry because of its theoretical advantages,
its superior ability to provide ``smooth" supernova light curves,
and accurate error estimates from first principles; no convolution or
resampling of any image data is involved.  Details of the technique are
given in the next section. The other approaches are mentioned here to
demonstrate that we made significant effort to determine the optimal
photometry for the supernova light curve analysis.


\subsection{Scene modelling photometry}

The main idea behind our scene modelling technique is to perform photometry
on individual calibrated images without degrading the PSF and without any
spatial resampling that leads to correlated noise between pixels. All
of the frames are fit simultaneously with a model of the galaxy
background plus supernova. This is statistically optimal in that the
model produces a prediction for each observed pixel that can be compared
to the observation and its error; propagation of pixel level errors to
fitted quantities is made in a precise and rigorous fashion.  

The basic concept is similar to the technique used by the Supernova
Legacy Survey (SNLS; Astier
\etal 2006), but developed independently, and includes a key new feature,
namely, no spatial resampling. The photometry described in  Astier
\etal (2006) is accomplished by modeling each image as the sum of a
time-independent galaxy background plus a time-dependent supernova and
convolving the model with a separate PSF for each image; however, all
images are resampled to a common pixel grid before doing the fit. This
leads to correlated errors between adjacent pixels, which, as described
by Astier \etal, lead to underestimated parameter uncertainties,
including the uncertainty on the supernova flux measurements. Astier
\etal estimate that the variances returned from the fit are 25\% too
small as a result of the pixel correlations. Because of this, they
adopt empirical uncertainties derived from multiple observations on a
given night.  They find that the typical variances as derived from repeat
observations are about 50\% larger than those predicted from the fits.
Our implementation does not involve any spatial resampling of the
images, so there are no correlated errors that can cause derived errors
to be underestimated. The tests described in Section \ref{sect:tests}
demonstrate that our error estimates are accurate.

We note that our technique provides the largest benefits when the
pre-supernovae template images are of comparable (or lower) S/N to the supernovae
images and/or the seeing in template images is worse than that of
the images with the supernovae. If high S/N and good seeing template
images are available, these can be resampled and degraded to the pointing
and resolution of the supernovae frames without introducing too much
correlated noise (because in this case the supernova frame, rather than
the template, dominates the noise). 

Aside from the modeling technique, our technique is customized for the
SDSS survey to take advantage of the pre-existing photometric catalog
of stars in the supernova fields.

We model each image as the sum of a set of stars, a
galaxy, a supernova, and background. The galaxy is modelled as a grid of
squares of constant surface brightness. The stars and supernova are
modelled as point sources, with magnitudes that are time-independent and
time-dependent, respectively. A separate PSF is determined for each image,
and each image is matched to the model convolved with the image's
PSF. A set of stars is used to determine the relative astrometric
and photometric transformations between the frames, and the stars and
supernovae are required to have the same relative positions in every frame.

The algorithm proceeds as follows. A set of calibration
stars is extracted from our calibration star list around the position
of each supernova. For each SDSS observation of each supernovae, a $2048
\times 1024$ pixel image subsection ($\sim 800 \times 400$ arcsec) is cut
out of the PHOTO corrected frames in each filter, with the supernova
centered in rows in the cutout (adjoining fields are pasted together if
necessary).  Hereafter, we
refer to the image subsections as frames. Since the SDSS data is taken
in drift-scanning mode, the mean time of observation differs by about 27
seconds from bottom to top of these image subsections.  The calibration
stars are sorted by brightness. Since the calibration stars only include
non-variable stars and do not extend to the faintest stars in the frame,
a star finding algorithm is used on a single $g$-band frame to obtain a more complete star
list. Using this star list, any object from the calibration star catalog
that has a nearby object is excluded, to ensure that the final calibration
list contains only isolated stars.

For each frame, a slowly varying background model is derived
by determining a sky value in 25 (a 5x5 grid) subsections within the
image. The sky level per pixel in each subregion is measured using an estimate
of the modal value (Stetson 1987) in the region; the final sky level
is remeasured after rejecting values $5\sigma$ larger than the initial
estimate (to further minimize effects of stars).  
A quadratic fit is done to these 25 values to provide a model
of the sky background. The rms of the 25 independent measurements is
compared with the standard deviation in the central region; if the
variation across the entire frame is larger than expected from the individual
variances, the frame is flagged as having a potentially uncertain 
sky level; the source of the variation in such frames can arise from
rapid changes in atmospheric conditions, and, in some cases, from
the presence of a very bright star in or near the image subsection.
Only a small fraction of all of our images show this behavior.

On each frame, stars are identified using the DAOPHOT (Stetson 1987)
FIND algorithm for potential use in determining the PSF; we use more
than just calibration stars for this purpose since even variable stars
are useful for PSF determination. This star list is filtered to remove
all objects with nearby neighbors, and any objects with shape parameters
that deviate significantly from those measured for the bulk of the stars.
On each frame, aperture photometry measurements are made for the stars in
the filtered list. A position independent PSF is created for the image
frame using all stars within 3 magnitudes of the brightest star in the
field.  A constant PSF gives an adequate representation (in most cases)
over the moderately small image subsection that we use. In any case,
there are generally too few stars to derive an accurate PSF model with spatial
variation. The PSF representation is made using the PSF characterization
of DAOPHOT (Stetson 1987): a Gaussian integrated over pixels is fit to
the brightest PSF star, and the residuals from this Gaussian are stored
in a lookup table at 0.5 pixel spacing. The removal of an underlying
Gaussian minimizes the effect of interpolation errors in the lookup
table. For any additional PSF stars, the integrated Gaussian from the
brightest star is fit to each star individually, and the residuals are
interpolated and added into the lookup table to reduce noise. The PSF
is assumed to be zero beyond a specified PSF radius.

We then proceed to fit a model to the observed data. At each pixel with
coordinates $(x,y)$ and in a given filter, the 
model for the flux is given by:
\begin{eqnarray}
M(x,y)  =  sky(x,y) + S \left ( \sum_{stars}I_{star} PSF(x-x_{star},y-y_{star}) + \right. \nonumber \\
I_{SN}PSF(x-x_{SN},y-y_{SN}) +  \nonumber \\
\left. \sum_{x_g,y_g}\mathcal{G}(x_g,y_g) PSF(x-x_g,y-y_g) \right) 
\end{eqnarray}
where $x$ and $y$ are the horizontal and vertical pixel indices,
$M(x,y)$ is the total model intensity (DN) at each pixel, 
$I_{star}$ is the known total calibrated brightness of each star, 
$I_{SN}$ is the unknown total calibrated supernova intensity, 
$PSF(\Delta X,\Delta Y)$ is the measured fraction of
light from a star as a function of the distance of each pixel from the
stellar position, $\mathcal{G}(x_g,y_g)$ represents the unknown grid of galaxy 
intensities,
and $sky(x,y)$ is the measured background value at each pixel.
$S$ is the unknown frame scaling factor that converts the calibrated
fluxes to DN on each individual frame. The positions ($x_{star},y_{star}$)
and ($x_{SN},y_{SN}$) are the pixel coordinates of the stars and
supernovae, which are derived from their celestial positions and
an astrometric solution for each frame.

The fits are weighted by the expected errors from photon statistics
and readout noise, using the gain ($G$; the number of DN per detected
photons) values for each camera column and
each filter as given in the SDSS {\tt fpAtlas} files. We adopted $\sigma_{rn}=5$
electrons for the readout noise; technically, the readout noise varies
from chip to chip, but a single typical value was adopted since it is 
a negligible noise source.  Specifically, we minimize:
\begin{equation}
  \chi^2 = \sum_{xy} {(O(x,y) - M(x,y))^2 \over (M(x,y)/G + ({\sigma_{rn}^2\over G^2}))}
\end{equation}
where $O(x,y)$ is the observed value at each pixel.
Operationally, we limit the model to include stellar (and SN) flux out
to a PSF radius of 5 arcsec from the center of each object. Due to lower
S/N in the outer regions of the PSF, only pixels within a specified
fitting radius (which is taken as 3 arcseconds or the measured FWHM of
the PSF, whichever is larger) are used in adjusting the fit parameters,
but the contribution of objects out to 5 arcsec is included in the model.

Since the model is nonlinear in the fit parameters, the solution is
iterated from a starting guess. Adjustments to the initial parameters
are computed using the full Hessian matrix, using a Levenberg-Marquardt
scheme.  If the fit has abnormally large $\chi^2$ after several
iterations, the weight of pixels with large \chisq
($>2.5\sigma$) is decreased; this attempts to prevent bad pixels from
corrupting the fit quality. The fit is judged to converge when all of the
point source intensities do not change significantly
in an iteration.

The first step in solving for the model parameters is to determine
accurate stellar positions for the stars on the calibration list. The
initial positions from the calibration catalog are average positions
from the pre-supernova template catalog. However, since the SDSS
template images of the SDSS supernova survey area go back to 2001,
proper motions are not negligible for some stars, and allowing for
proper motions significantly improves the quality of the model fits to
the data. Our first fit solves for stellar positions and proper motions
using a subset of the SDSS $r$-band images. For this fit, we take the
initial epoch and subsequent epochs separated by 60 or more days from
the previous epoch. To maximize the baseline for proper motion determination,
we use all SDSS data taken from the beginning of SDSS survey (2001) until
the end of the SDSS SN survey (2007); this typically gives us 10-20 images to fit. The
stack of image subsections is simultaneously fit for stellar positions
(at epoch 2000), proper motions, an astrometric solution for each frame,
and photometric frame scalings between the frames. We arbitrarily adopt
the SDSS astrometric solution of the first frame in the list as the
absolute reference frame, since all we really care about is accurate
relative astrometry between the frames. This process yields us a list
of stars with accurate relative positions on the sky, proper motions,
and calibrated brightnesses. Since the fit only includes stars, the
proper motions are not absolute, but are only relative (in the fit, we lock the
proper motion of the first star to be zero); after the fit, we
normalize them so that the mean proper motion of all of the stars is zero
(but we allow for a proper motion of the reference frame in the galaxy
fit, see below).

For the astrometric model, we adopt the distortion coefficients measured
by the SDSS photometric pipeline, but solve for a full linear astrometric
solution (6 parameters) within each of our subframes. For any frames
where only 3 calibration reference stars are available, we constrain
the astrometric solution to fit only 4 parameters for scale, rotation,
and offset.

Given measured stellar positions, our second series of fits solves for the
frame scale factors, $S_{frame}$, and the astrometric parameters for each
frame. These can be determined for each frame independently, since all of
the stellar parameters (positions, proper motions, and intensities) are
held fixed in the fit. Only frame parameters (which are independent from
frame to frame) are solved for; in these fits, there are 7 parameters
(6 linear astrometric parameters plus 1 photometric frame scaling).
A single photometric frame scaling value for our subsections requires stable
transparency over a time interval of $\sim$ 81 seconds, and over a spatial
scale of $\sim$ 800 arcsec.  Based on the residuals of stars across the
field, we have found, to no surprise, that the assumption of a single photometric frame
scale value becomes less accurate under cloudier conditions. As a result,
we flag all frames where the photometric scaling is less than half the
expected scaling for photometric weather (allowing for differences in
airmass). 

To identify frames that may have other problems, and to assess the
quality of the astrometric/photometric solution, a final fit iteration
is performed after the frame solution is determined; in this final
iteration, we lock the frame parameters and stellar positions and fit
for the individual stellar brightnesses. These recovered brightnesses
are compared with the known brightnesses from the calibration star
catalog. A subset of the best measured stars is selected so that it
contains at least 5 stars (3 in the $u$-band). The mean magnitude
difference, rms, and \chisq for this set of stars are computed using
the fit brightnesses and error estimates; for the \chisq calculation, an
error term is included for the uncertainites in the calibration magnitude
of each star. The reduced \chisq  for the frame is recorded, and all
frames with atypically large \chisq are flagged. Finally, we estimate a
``frame error'' term by determining what additional error needs to be
added (in quadrature) to bring the reduced \chisq down to unity; this term is generally
less than 0.01 mag, and is plausibly associated with errors that result
from inaccuracies in the PSF model.

\begin{figure}
\plotone{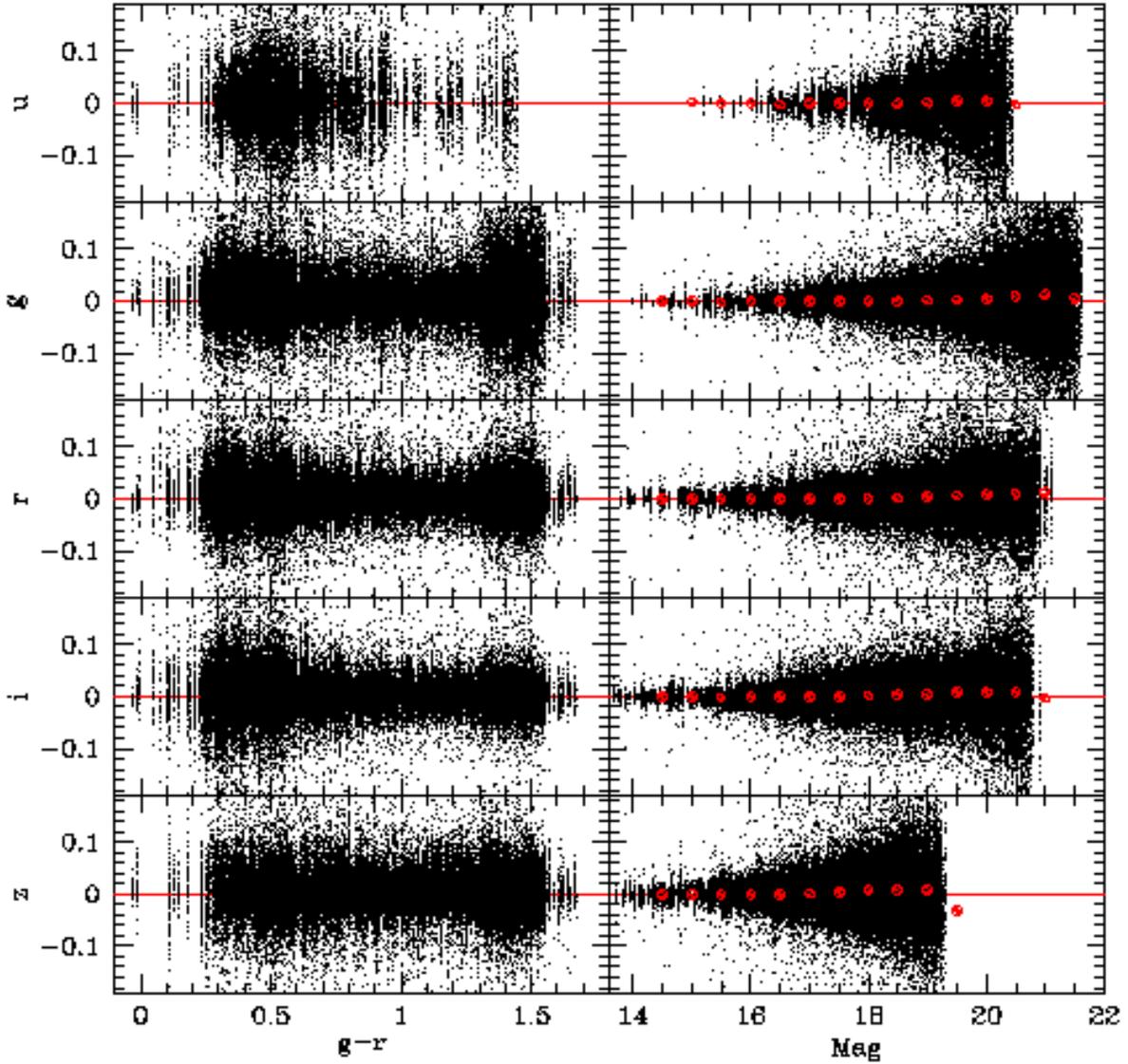}
\caption{Difference between recovered stellar magnitudes and
the calibration magnitudes as a function of stellar color (left
panel) and stellar magnitude (right panel). This demonstrates
the accuracy of our PSF photometry and also indicates typical errors
as a function of stellar brightness.}
\label{fig:sdssstars}
\end{figure}

Figure \ref{fig:sdssstars} shows the difference of the recovered stellar
magnitudes and the calibration magnitude for all of the calibration
stars for all of the 2005 confirmed type Ia supernovae as a function of
stellar brightness and color. These plots demonstrate that we accurately
recover the brightnesses of the calibration stars with our PSF fitting,
and display the typical photometric errors in our exposures as a function
of stellar magnitude.

The derivation of astrometric parameters and photometric scaling factors
for each of the supernovae frames discussed so far is similar to what is done 
for most supernovae surveys, although the inclusion of proper motions
may not be typical (or needed, when the time baseline is short). We also
have attempted to do a careful accounting of errors.

In the third, and final, fitting stage, we extract a small $128 \times 128$ 
($\sim 50 \times 50$ arcsec) image subsection around the position
of the supernova in every frame. Using the derived frame photometric
scalings and astrometry, we simultaneously fit the entire stack of images
(all epochs and all filters) to solve for a temporally constant galaxy
plus a temporally variable supernova. Frames that have been flagged as
potentially unreliable in any of the previous steps are not allowed to
influence the galaxy model, but are still included for a determination
of the supernova brightness (as described below, the flag is carried
along for the final output). We obtain the estimated supernova peak
intensity date from the search photometry, and force all observations
more than 90 days before peak to have zero supernova flux. For a typical
supernova, the image stack contains several hundred images: 10-20 pre-SN
images and 10-20 SN images in each of 5 filters. A single supernova
position is fit to the entire stack.  The galaxy is modelled as a grid
of squares of constant surface brightness; we use a 15 by 15 grid of
0.6 by 0.6 arcsec squares, with independent brightnesses in each of
the 5 filters at each location. The model galaxy size of 9 by 9 arcsec
around the position of the supernova may not model the entire galaxy,
but models a sufficient amount to determine the galaxy contribution at
the position of the supernova even under the worst seeing conditions.
The galaxy model is interpolated to the pixels on each frame separately;
the choice of the model grid spacing is not critical.  Given the typical
SDSS seeing of $\sim$ 1.2 arcsec, it is clear that the information at
the 0.6 arcsec scale is limited, and in fact, the recovered galaxy
maps often do not show realistic structure at this spatial scale. However,
when the recovered maps are smoothed to the typical seeing, they match
the observed galaxy well, and the relatively fine sampling allows us
to match regions with steep intensity gradients.  We have investigated
using both coarser and finer samplings for the galaxy model, and find
that the supernova photometry is relatively insensitive to sampling changes.  
The supernova is allowed to have a separate brightness
in each frame, but is required to have a common position in \textit{all}
frames; the position is iteratively determined by the fit using all of
the available data. The total number of fit parameters in the final fit
is \begin{equation} N_{fit} = (15\times 15) N_{filt} + N_{epoch}*N_{filt}
+ 4 \end{equation} where $N_{filt}$ is the number of filters (usually 5,
but sometimes 4 if there are an insufficent number of $u$-band calibration
stars) and $N_{epoch}$ is the number of epochs observed later than 90 days
before the estimated SN peak. The final four parameters are for the
celestial position ($\alpha, \delta$) of the supernova, and the mean proper
motion of all of the calibrating stars in the field; it is the galaxy
light that constrains the mean proper motion of the calibration stars.
\footnote{For hostless supernovae, the mean proper motion would be 
unconstrained, but for such objects, there is no pre-SN galaxy
background that needs to be accounted for, and proper motions are negligible
over the decay time of the supernova.}

Output from the final fit includes supernova brightnesses for each frame
along with error estimates from the least-squares fit.  Since the noise
model is derived from photon statistics and readout noise, but does
not include terms from an imperfect PSF, inaccuracies in the determination
of the frame photometric zeropoints, or sky model, the least-squares
errors may be underestimated, especially for the brightest points where
statistical errors are small. In an attempt to provide realistic error
estimates for all points, we take the errors from the fit and add in
the individual ``frame errors'', the derivation of which was previously
described. Since these are derived from observations of relatively bright
stars, they are expected to account for errors in PSF modelling and 
frame scaling.

Two other sources of error are also considered: error arising from
inaccuracies in the sky estimate and error from inaccuracies in the galaxy
model at the location of the supernovae. The former gives a systematic
error over the pixels covered by a supernova at any individual epoch,
but is likely to be a random error source for different supernovae
epochs. The galaxy model error gives systematic errors that are similar
(not identical, because of seeing variation) for all epochs of a given
supernova. We estimate the sky error based on the variation of sky level
from different subsections of the frame (although in cases where there
is real structure in the sky background, this might overestimate the
sky error).  The galaxy error is calculated from the least squares fit,
and includes correlated errors which exist between adjacent locations in
the galaxy model, since this model is sampled finer than the point spread
function; this estimate of the galaxy error may be an underestimate since
it does not account for errors that would result from systematic errors
in the astrometric solution of the frames.  Since the portion of the
galaxy that contributes flux at the location of the supernova depends on
the seeing, the galaxy error can vary from frame to frame; for output, we
calculate a typical galaxy error that arises for a seeing of 1.2 arcsec.
The estimated errors from inaccuracies in sky and galaxy subtraction
are output, along with the supernova brightness and its random error.
Clearly, the importance of the sky and galaxy subtraction errors is
larger when the supernova brightness is comparable or fainter than the
sky or galaxy. In general, errors in the sky background dominate those
in the galaxy model.


We have created images of the frame with the model subtracted, and
inspection of these provide qualitative confirmation the quality of the model
(see below for some quantitative tests).

\begin{figure}
\plottwo{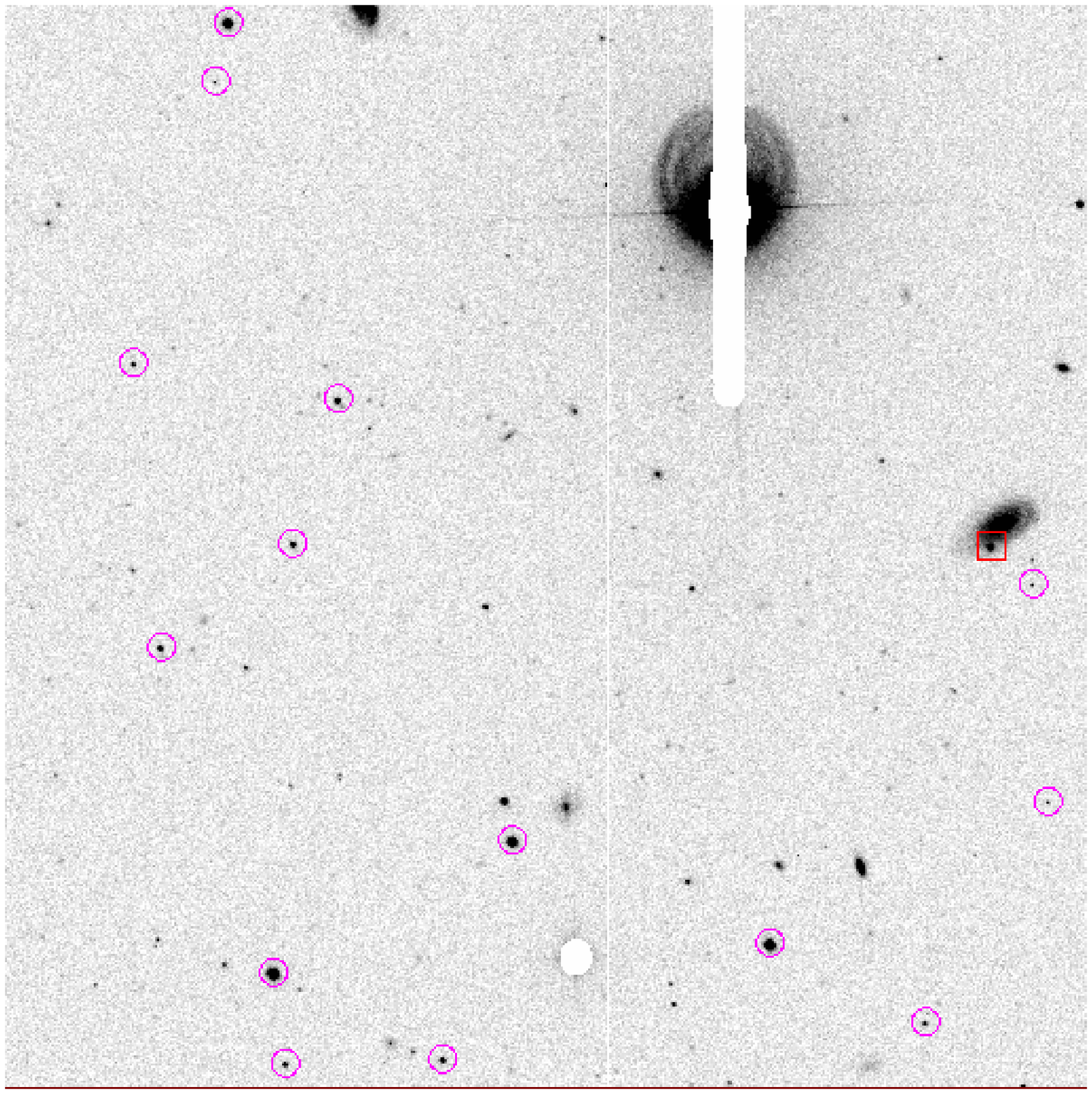}{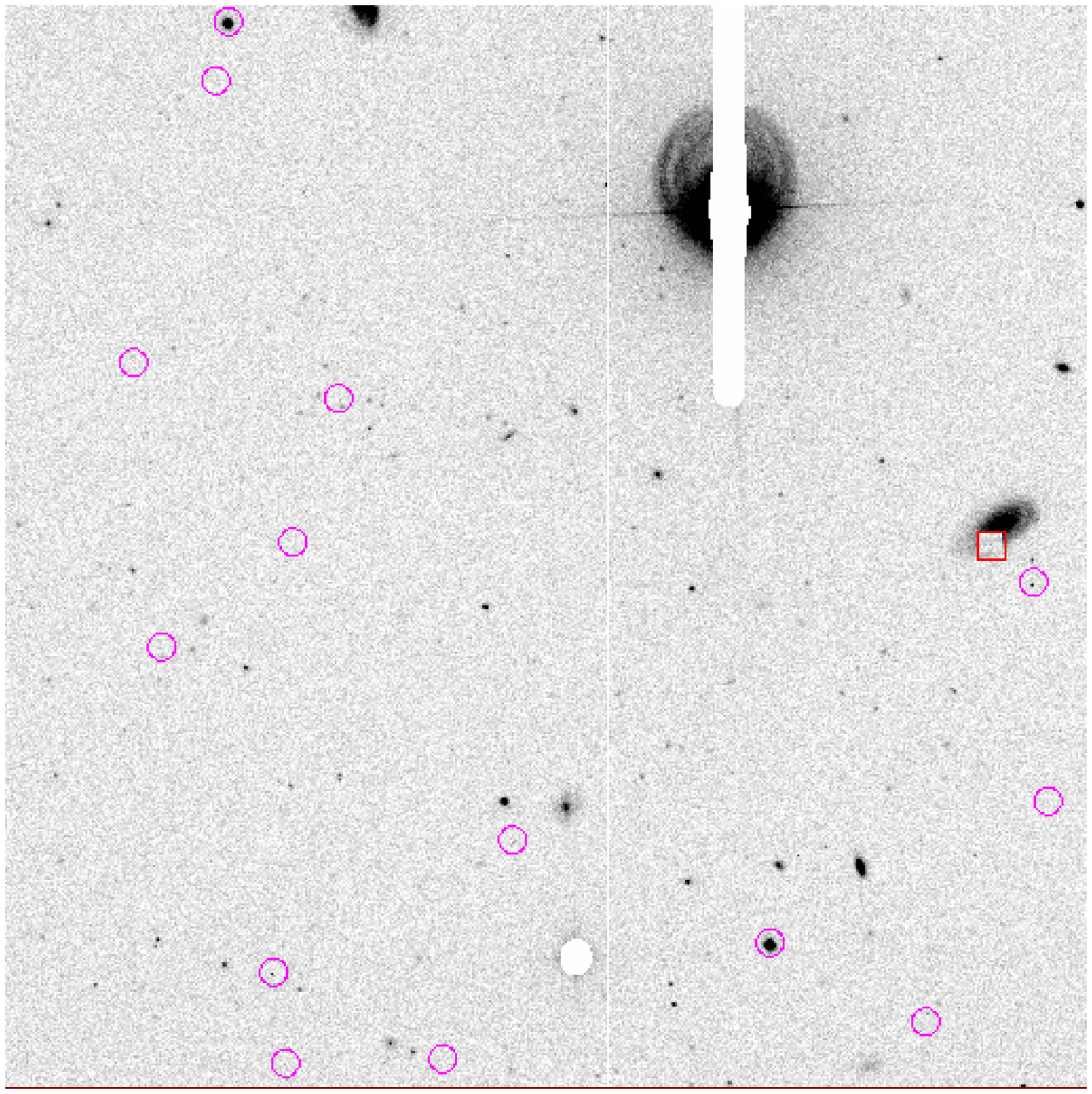}
\caption{An example of an image subsection used to solve for frame parameters
in the second fitting stage, \ie, the astrometric solution and 
photometric scale factor. The stars with
circles are the calibration stars used to determine the solution.
The left panel shows the image before the model is subtracted, with 
circles around the calibration stars and a box around the supernova;
the right panel shows the image after model subtraction.}
\label{fig:sample}
\end{figure}

\begin{figure}
\plottwo{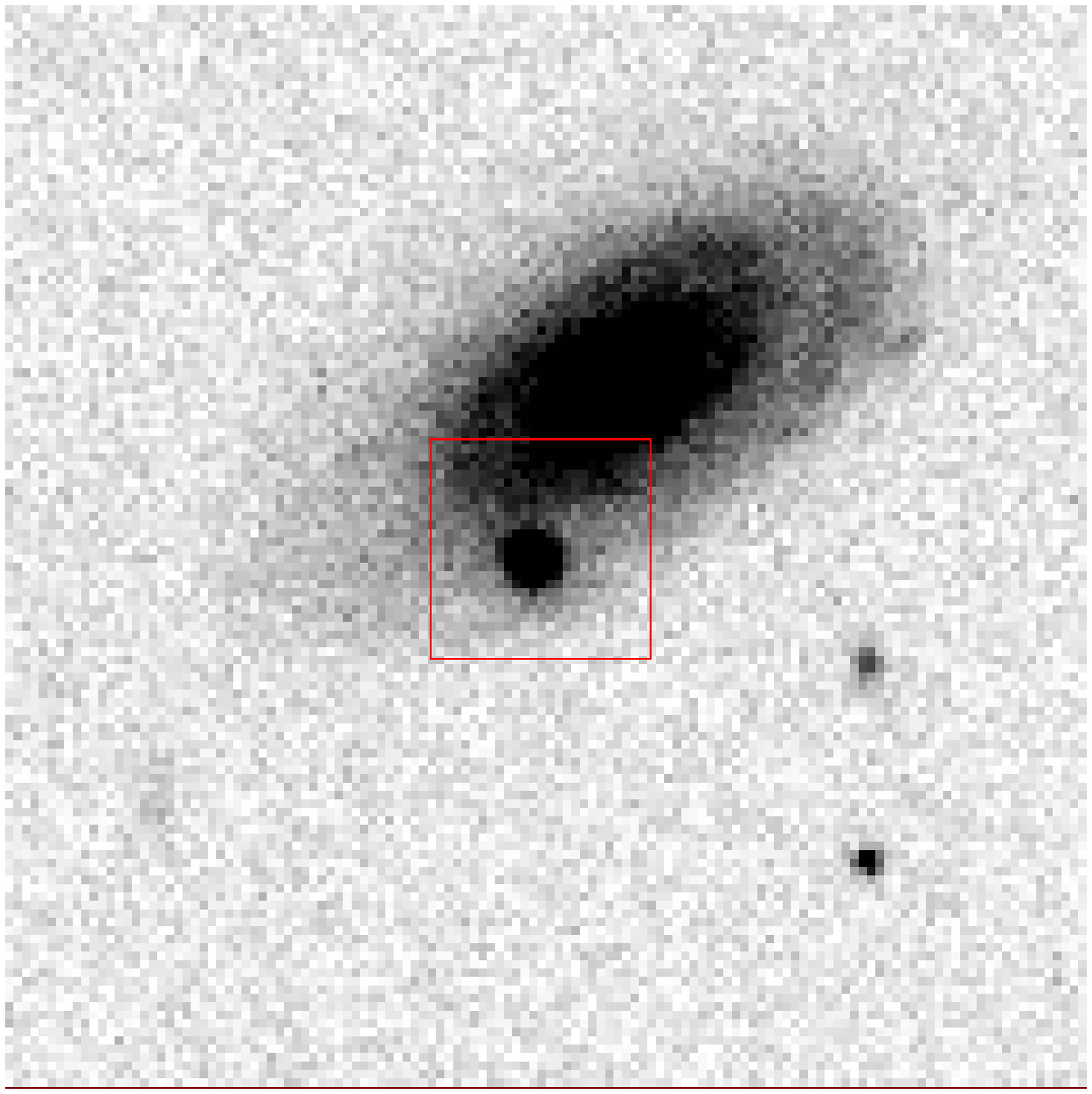}{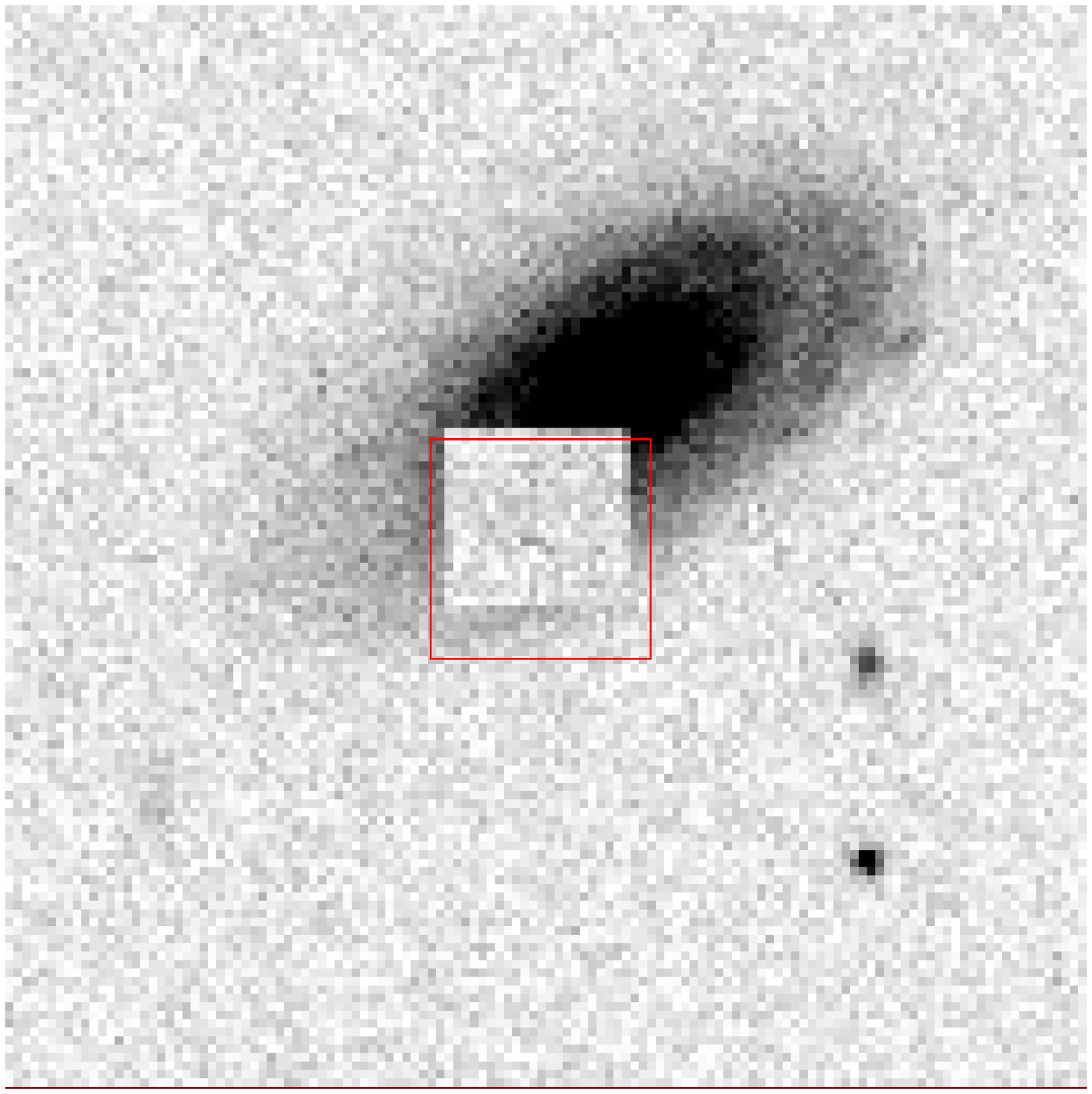}
\caption{An example of an image subsection used to solve for galaxy background
and supernova brightness. In this third fitting stage, an entire stack
of images, including those with and without the supernovae present,
are fit simultaneously.  The left panel shows the image before the model is
subtracted, with a box around the supernova; the right panel shows the image 
after model subtraction.}
\label{fig:samplez}
\end{figure}

Figures \ref{fig:sample} and \ref{fig:samplez} show an example of
the procedure applied to one measurement of one of our supernovae.
Figure \ref{fig:sample} demonstrates the initial astrometric and
photometric solution that is determined for each frame individually.
The left image shows the image subsections that is used; circles
show the calibration reference stars, and the square shows the
supernova. The right image shows the same frame after the best fit model
has been subtracted.  Figure \ref{fig:samplez} shows the region around
the supernova that is used to simultaneously solve for galaxy background,
supernova position, and supernova brightness at each epoch; in this stage,
an entire stack of these images is fit simultaneously.

We note that the scene modelling technique does not require perfect
spatial overlap between all of the images, so long as there are some
stars in common in all of the frames to allow
determination of accurate relative astrometry.  For supernova that lie
in the overlap between strips 82N and 82S, there may be few, if any,
reference stars in common between the northern and southern strips. In
these cases, the entire dataset is still fit simultaneously. However,
the final iteration allows for a global shift between all of the frames
in one strip to those in the other strip; similarly, the two strips are
allowed to have different mean stellar proper motions.  It is the galaxy
itself that provides the information to determine the global shift and
proper motion difference between the two strips.

\subsubsection{Data flags}

\label{sect:flags}

For each supernova measurement, we set a flag to allow for
points of potentially poorer quality to be recognized. The value of the flag
is a bitwise combination of multiple criteria:
\begin{itemize}
 \item[1] Sky brightnesses more than twice median sky brightness 
   from entire stack of images in this filter, \ie, moon or clouds
 \item[2] FWHM of stellar images larger than 2 arcsec, \ie, poor seeing
 \item[4] photometric scale factor less than 0.5, \ie, moderately
   cloudy conditions
 \item[8] atypical sky variation: ratio of sky variation 
   between image subsections to sky variation within a subsection
   significantly larger than median of entire stack in this filter
   (can arise in cloudy conditions)
 \item[16] large sky variation: ratio of sky variation between image 
   subsections large compared with sky variation within a subsection
   (can arise from presence of bright star nearby)
 \item[32] derived supernova brightness fainter than underlying
   galaxy brightness (measured using the PSF of the frame)
 \item[64] Fewer than 5 calibration stars on frame
 \item[128] rms photometry of calibration stars atypically large
 \item[256] fit exceeded maximum number of iterations, or fit quality
   (from individual frame \chisq) poorer than typical
 \item[512] No calibration stars 
 \item[1024] Photometric scale factor so low, rms photometry of calibration stars 
   so high, variation in sky brightness so high, frame fit quality so poor, or
   global fit quality so poor, to strongly suggest that data should not be used, 
   i.e. \textbf{bad data}.
\end{itemize}

The highest quality points have a flag value of 0. Flag bits of
1,2,3,8,16,64, and 128 are determined from the individual frame fits
before the global galaxy/supernova solution is determined.  Frames with
any of these flags set are not used to influence the galaxy model in
the final fit, with the exception of flag=16.  This flag can be set
because of background light from a nearby very bright star.  In this
case, the problem persists at all epochs, and a result can be obtained
only if these frames are used.

Most points with $0<flag<1024$ appear to be of good
quality judging from how well they fit on the light curves.  Bit 6 (32)
flags points where the supernova is fainter than the underlying galaxy,
and as a result, applies to many points for objects buried within bright
hosts and to many late time points. These are the points that are most
sensitive to the accuracy of the galaxy model, and are most subject to
the possibility of systematic error.

Observations with the 1024-bit set, i.e.  $flag>1024$ are generally unusable, 
and should not be trusted. For applications where only the cleanest (potentially
highest accuracy) data are desired, even at the expense of throwing away
many apparently good points, one might choose to only use points with $flag=0$.
The SDSS cosmology analysis (Kessler \etal 2009) uses essentially all 
points with $flag<1024$.

%


\subsubsection{Including non-2.5m data in scene modelling}

\label{sect:non25m}

An important feature of the scene modeling technique is that the model
is independent of telescope characteristics such as pixel size and
registration relative to the model.  It is therefore straightforward
to combine data from different telescopes in the same fit.  The same
catalog stars can be used to calibrate the response of all the telescopes.


In general, each telescope will have its own unique set of filter
response curves. As a result, relative photometry of objects with
different spectral energy distributions will differ from telescope
to telescope. If the differences in filter response from telescope to
telescope are small, then the differences can be parameterized by use of
a linear color term. For the non-2.5m data, when deriving the photometric
scaling for each frame from the calibration stars, we allow for a color
term to be fit as well as a photometric zeropoint.  Since we expect the
color term to be constant in time, at least over an observing season,
we adopt an average color term from the photometric solutions for all
frames in a given filter using all of the supernovae observed in the 2005
season; this allows for a large range of stellar colors to be sampled.

For the 2005 season, photometry of the SDSS supernovae was obtained with
several other telescopes; in most cases, only 4-color (\textit{griz}) 
observations were obtained.  Each individual frame was fit to the calibration
star list derived from the SDSS frames exactly as above, except a color
term was included when fitting the instrumental brightnesses to the
catalog brightnesses. For each telescope, color terms of the form
\begin{eqnarray}
g = g_{obs} + t_g (g-r) \\
r = r_{obs} + t_r (r-i)\\
i = i_{obs} + t_i (r-i) \\
z = z_{obs} + t_z (i-z) 
\end{eqnarray}
were determined, requiring time-independent transformation coefficients over the
length of the observing season.

\begin{figure}
\plotone{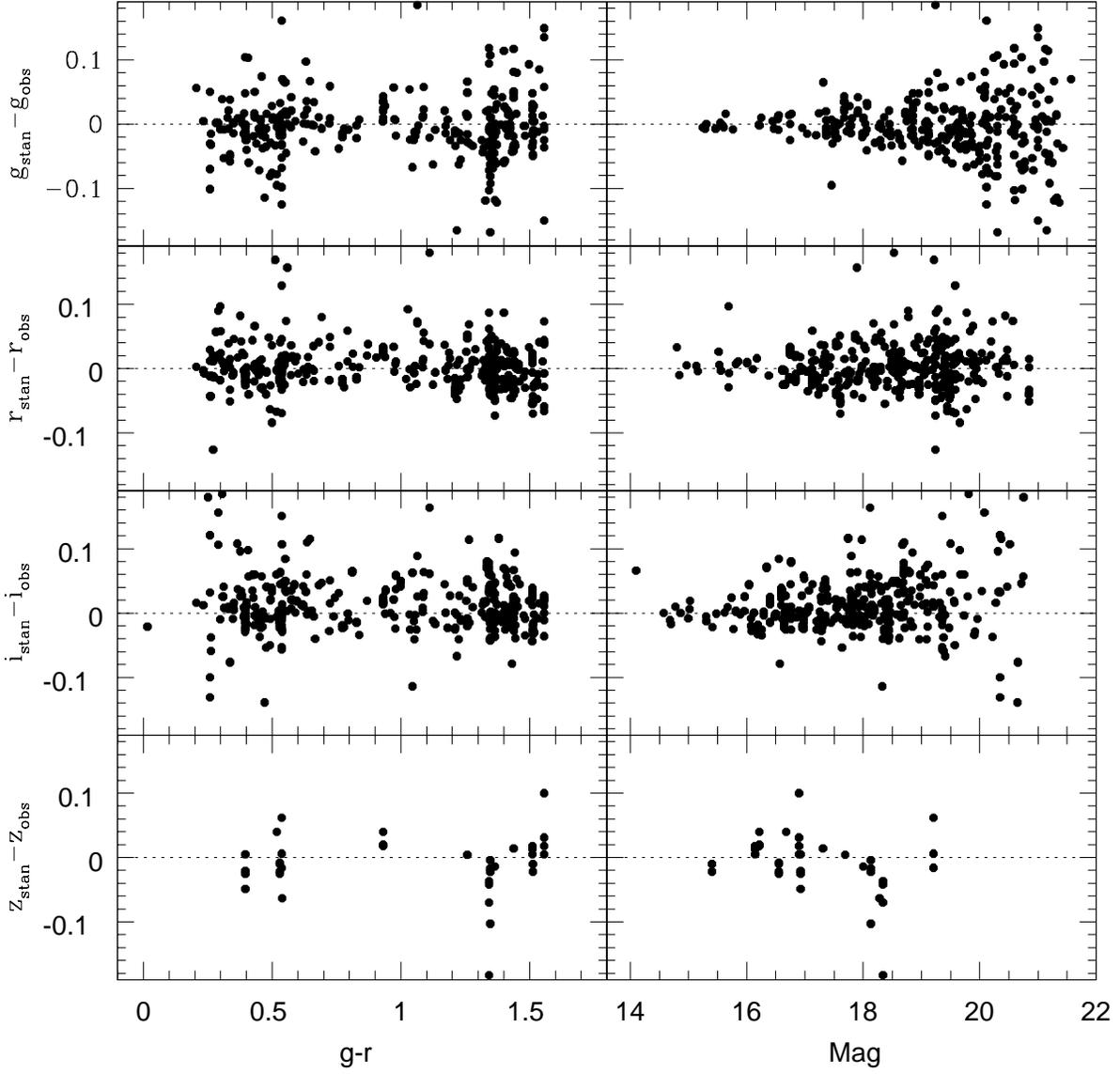}
\caption{Difference between recovered and calibration photometry for 
calibration stars in the MDM frames, using the color terms presented in 
the text.}
\label{fig:mdm24m}
\end{figure}

Figure \ref{fig:mdm24m} shows an example of the photometric calibration 
results for all of the stars on MDM 2.4m frames from the 2005 season after the
derived color terms have been removed;
this plot is equivalent to Figure \ref{fig:sdssstars} for the SDSS frames.
The adopted color equations (that are applied for this plot) are:
\begin{eqnarray}
g = g_{mdm} - 0.1 (g-r) \\
r = r_{mdm} - 0.05 (r-i)\\
i = i_{mdm} + 0.08 (r-i) \\
z = z_{mdm}
\end{eqnarray}

Similar relations have been derived for the other telescopes used during
the survey.


The differing filter responses also affect the underlying galaxy background.
To account for this, we apply the stellar color term to the underlying
galaxy model as well. The accuracy of the application of a color term
depends on the degree to which the spectral energy distribution of the
object to which the color term is applied (the galaxy, in this case) is
similar to the SED of the objects (stars, in this case) used to derive
the color term. While SEDs of galaxies are not identical to those of stars,
at the moderate redshifts considered here, they are not dramatically
different. Combined with the fact that the color terms are relatively
small (since SDSS-like filters were used on all of the non-2.5m telescope),
we feel confident that the application of the stellar color terms
to the galaxy background model is adequate. 

For the final supernova photometry, the non-2.5m frames can be included in
the final photometry iteration described above. However, to ensure that
any issues with the photometric transformation for the non-2.5m data
do not deteriorate the quality of the SDSS 2.5m photometry, we do not
allow the non-2.5m data to contribute to the solution of the galaxy
model itself; only 2.5m data is used to constrain this model, and the
inclusion (or lack thereof) of non-2.5m data has no effect on the 2.5m 
photometry.

Interpreting the supernova photometry from the non-2.5m data can
be challenging, because supernovae have spectral energy distributions that are
quite different from stars. As a result, application of color
terms derived from stars does not necessarily bring supernova photometry
onto the 2.5m system. 
Clearly, the use of these data in conjunction with
the 2.5m photometry requires some understanding of the response
differences between the telescopes and the spectral energy distribution
of the supernovae at different epochs (\eg, via so-called S-corrections).

Unfortunately, it is usually rather difficult to get accurate measurements
of the response functions of different systems.  For some of the
telescopes we have obtained synthetic response functions from
combinations of response functions of individual components.  However,
color terms computed from application of these response functions with
libraries of stellar spectral energy distributions do not always match
the measured color terms, suggesting errors in the response functions or
the stellar libraries. This suggests that extreme caution should be used
when applying products of individual component responses to determining
transformations between observations using different photometric systems.
We plan to investigate this in detail using measured response curves, the
stellar calibration data, and several near-simultaneous observations of
supernovae by multiple telescopes.

Since the weather at APO was quite good for the 2005 season, the
2.5m light curves provide good coverage even without the non-2.5m data.
Because of this and the complication of understanding the system
responses of the non-2.5m data, we have chosen not to include these
data in our initial analyses, and in the data release described in this
paper. However, we hope to do so in the future,
especially since we expect the other telescopes to contribute more in
the last two observing seasons, mostly through followup of objects
discovered late in the 2.5m observing season; in 2005, many of the late
objects do not have sufficient coverage to make them useful. It is in
anticipation of using these data that we have included the discusion of
the application of scene modelling to non-2.5m data here.


\section{Photometry tests}

\label{sect:tests}

We have performed a number of exercises to verify and improve the quality
of the scene modelling photometry and error estimates. These tests also
allow us to make educated decisions about what, if any, data cuts should
be made before light curve analysis.

\subsection{Stellar photometry}

The first test treats real stars nearby the supernovae as if they were
supernovae themselves, recovers light curves for them, and compares the derived
brightnesses with the standard star catalog brightnesses. Using
stars near the supernova allows us to accurately understand how any
errors in astrometry and the PSF are likely to affect the supernova
photometry. Unlike supernovae, there is no galaxy background
underneath these stars, but fitting for a model background that is zero is
a valid, if somewhat unrealistic, test. In order to simulate an underlying
zero galaxy background, we remove the star from the early epoch frames by
replacing it with sky background taken from a nearby region of blank sky.
The full stack of images (including the early epochs
with star removed, and later epochs with star retained) was then run
through the scene modelling software, allowing for a background to be fit.

Note, however, that the ``known'' calibration magnitudes (taken from
Ivezic et al. 2007) are actually not \textit{perfectly} known, and any
errors in these will lead to increased scatter in our comparison (which
includes many stars). To compensate for this, we have averaged all of
our measurements of these stars (which make for many more measurements
than went into the Ivezic et al. catalog!), and compare the individual
measurements against this refined average.

\begin{figure}
\plotone{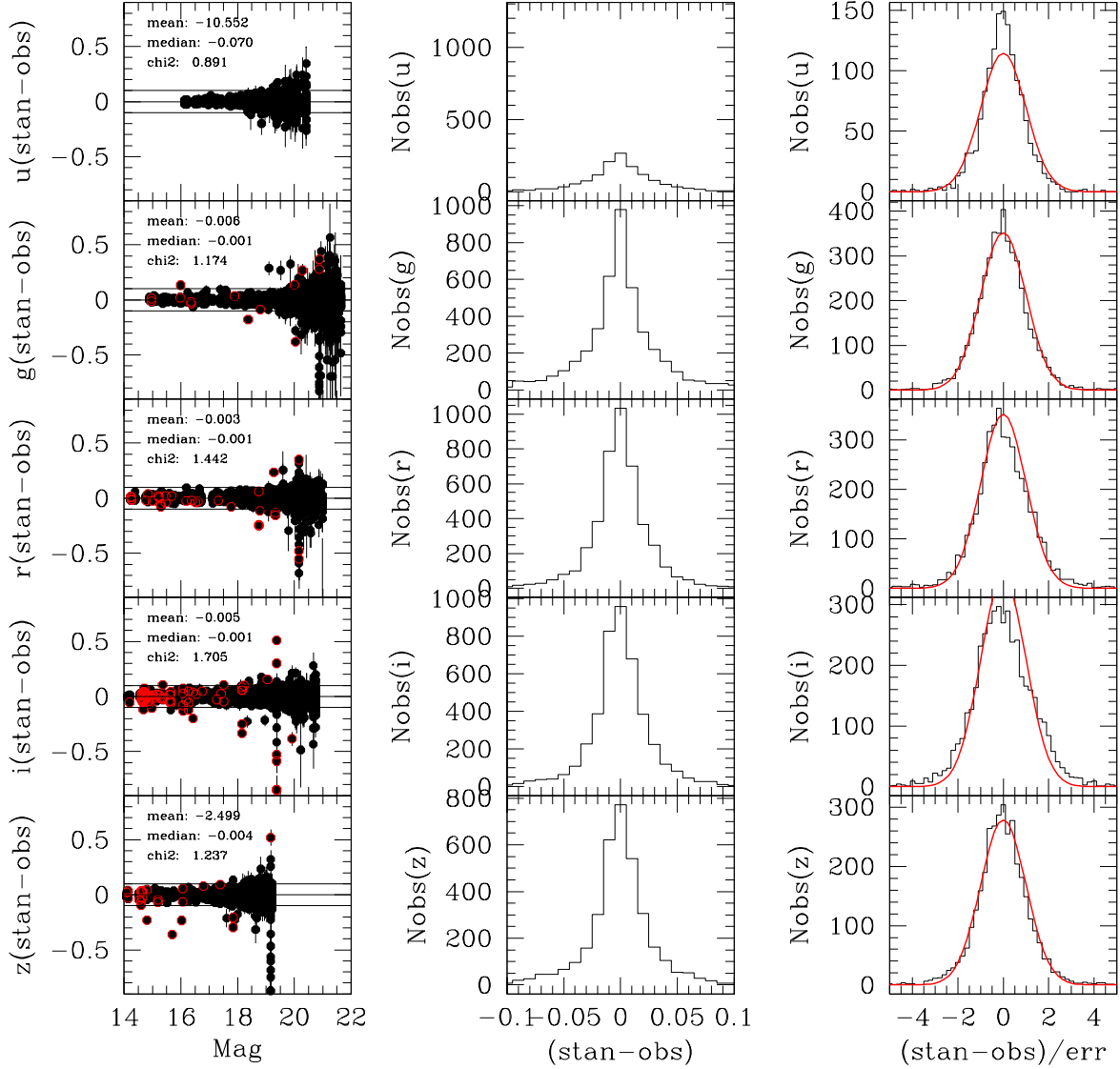}
\caption{Photometry of stars near the 2005 SN, treating them as if
they were supernovae, allowing for an underlying background
to be fit. The left panel shows the difference between the recovered
magnitude and the known stellar magnitude a function of magnitude.  
The central panel shows a histogram of error in the recovered magnitude,
and the right panel gives a histogram of difference between recovered 
and calibration magnitudes, normalized by predicted photometric error. 
The curve in the right panel
shows the expected Gaussian for the difference if the calculated errors are 
correct.}
\label{fig:stars2}
\end{figure}

Results are shown in Figure \ref{fig:stars2}. The left panel plots
the error in recovered brightnesses as a function of the stellar
brightness.  The central panel gives a histogram of the difference
(standard - observed).  In general the recovered brightnesses are,
to within estimated errors, consistent with the known brightness, with
median errors from the entire sample of stars of only a few millimag.

\begin{figure}
\plotone{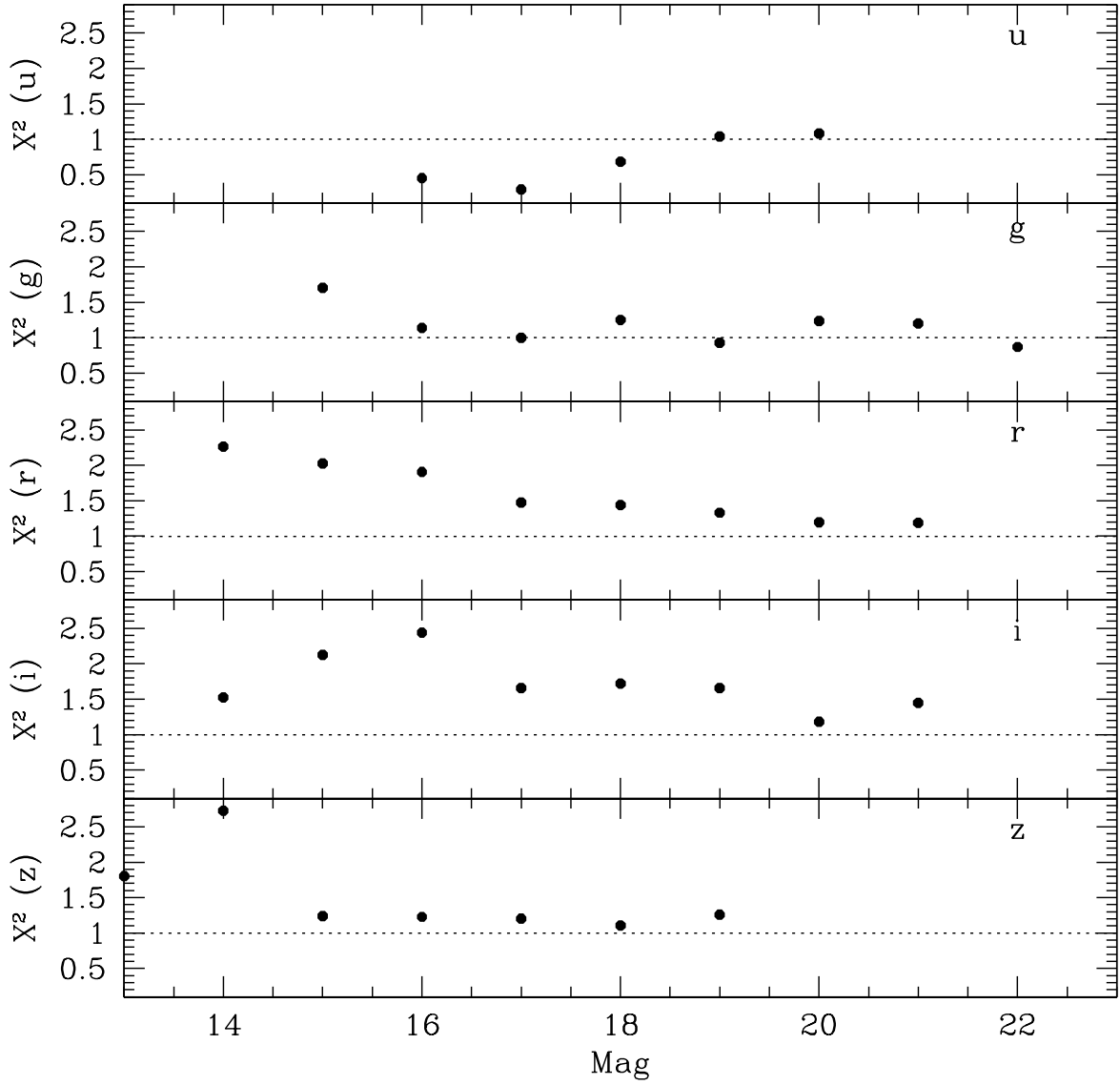}
\caption{Reduced $\chi^2$ for stars measured as if they were supernovae, 
computed by comparing the individual recovered magnitude against their
known magnitude. No sky error term has been included. }
\label{fig:starsum}
\end{figure}

The right panel shows the histogram of the difference normalized by
the calculated error; if the error estimates were perfect, this should
be a Gaussian of unit width. A more quantitative analysis of the error
estimates is given in Figure \ref{fig:starsum}, which shows the calculated
reduced \chisq from these distributions. For the brighter stars, the
reduced \chisq are near the expected value of unity for some filters, but
they are a bit too large for other filters. We have investigated the
source of this, and find that the larger \chisq come mostly from
points with small predicted errors, less than 0.01 mag. This suggests
that even with our procedure of adding a frame error, we still slightly
underestimate our errors for the brightest sources.  If we were to
impose a 0.01 mag floor on the predicted errors, the \chisq for
the stars comes down near unity in most cases. We note that our
supernovae are essentially never so bright as to have such a small
error. For the fainter stars, the \chisq values are slightly too large
in the $r$ and $i$ filters. This likely arises because no sky error has
been included in these error estimates (see section \ref{sect:fakes}).




\subsection{Pre-supernovae measurements}

To test for errors in modelling a real underlying galaxy background,
we measured the supernovae flux for real supernovae at epochs before
the supernova actually occurred, to see how well we would recover zero
flux. Clearly, the galaxy model depends on the pre-SN epochs and the
quality of that model will deteriorate if we remove too many of the pre-SN
epochs from the list of images with constrained zero supernovae flux.
Because of this, we chose to do this test using the 2005 data, but
looking at locations where supernovae were discovered in 2006. This
provides a good representation of the real situation for the 2005 supernovae.

\begin{figure}
\plotone{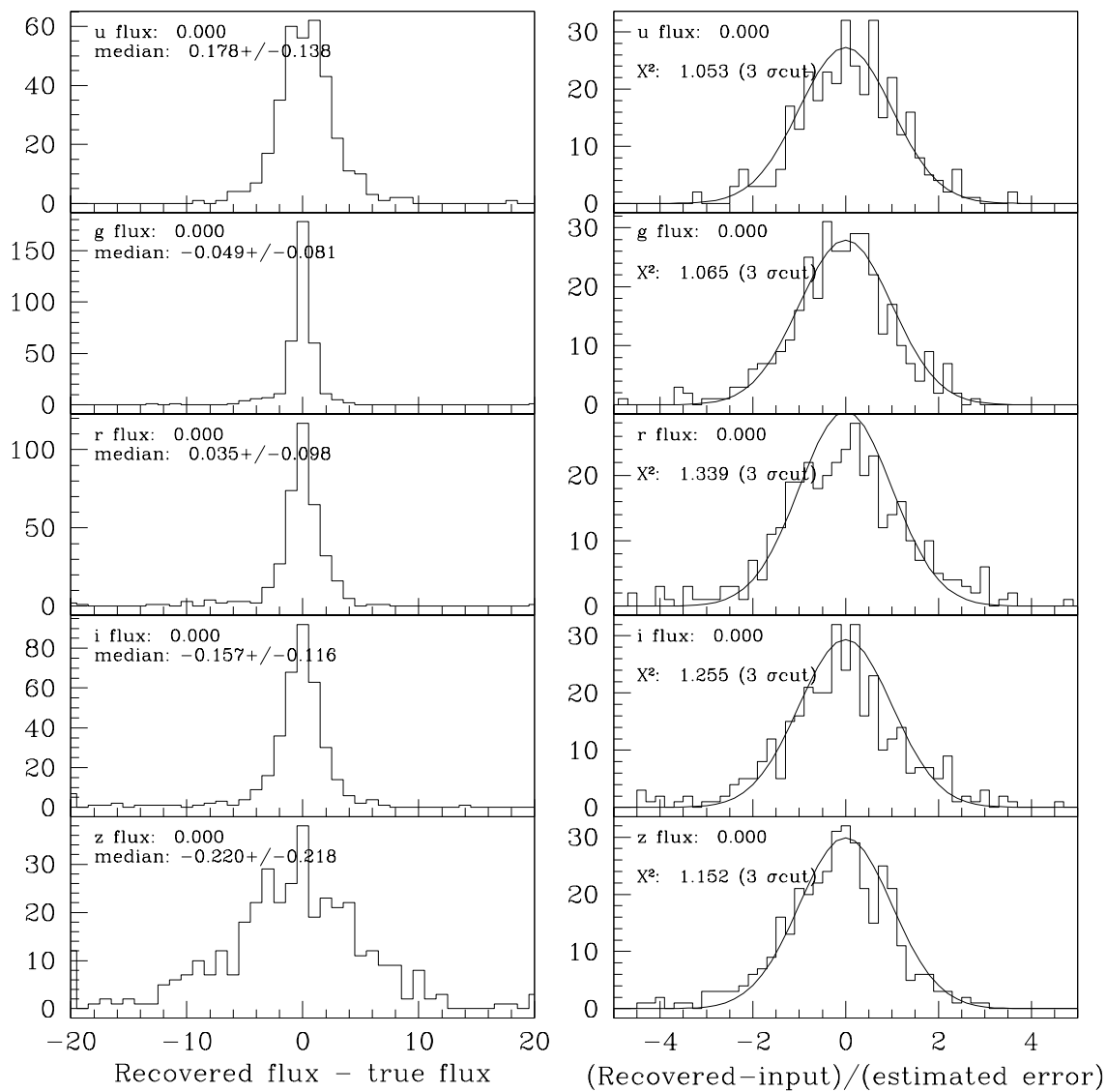}
\caption{Photometry of 2005 epochs at location of 2006 supernovae,
treating these as if they could have SN flux. Left panel shows histogram
of recovered fluxes, which should be zero; the units are $\mu Jy$. 
The right panel shows a histogram of recovered flux
normalized by predicted photometric error. The curve 
shows the ideal Gaussian of unit width.}
\label{fig:fake0}
\end{figure}

Figure \ref{fig:fake0} shows the results for measurements at the
location of the 2006 supernovae in the 2005 data.  The ideal situation
is to measure identically zero supernova flux. The left panel shows the
histogram of the difference between the observed and zero flux, in
units of microJansky ($1 \mu Jy = 10^{-29} ergs/cm^2/s/Hz$; a source
with apparent magnitude of 20 has a flux of 36.31 $\mu Jy$).  
The right panel shows the histogram of the magnitude difference normalized
by the predicted error. While there are a few points with measured
brightness significantly different from zero, the bulk of the distribution
follow the expected normal distribution. A more quantitative
discussion of the estimated errors in presented in the following section.

\subsection{Artificial supernovae}

\label{sect:fakes}

Finally, to test the accuracy of photometry at low flux levels against
the galaxy background, we inserted artificial point sources into
the frames and measured their brightnesses. Again, we used the locations
of the 2006 supernovae to place artificial sources into the 2005 frames.
We inserted artificial supernovae at 11 different flux levels. To
reduce computation time, we split the 2006 sample of about 250 supernovae
locations into 11 groups, so at each level, we inserted artificial
objects about 22 different sky locations; at each location, artificial
objects were placed in about 20 different epochs in the 2005 observations.

Artificial supernovae tests are not perfectly realistic because one
must assume an astrometric solution, a photometric solution, and a PSF
to insert the artificial objects, and usually the same quantities
are used in the data reduction. For situations where uncertainties in
any of these are the dominant source of error (bright objects), artificial supernovae
tests are likely to provide overly optimistic results. As a result, we
performed these tests only at a range of low flux levels.
Artificial supernovae were placed into the frames using the derived
astrometric solution and photometric scalings, and the measured PSF. The entire
stack of images was then run through the scene modelling software, and
the measurements at the supernova position were compared with the known
artificial supernovae brightnesses.

\begin{figure}
\plotone{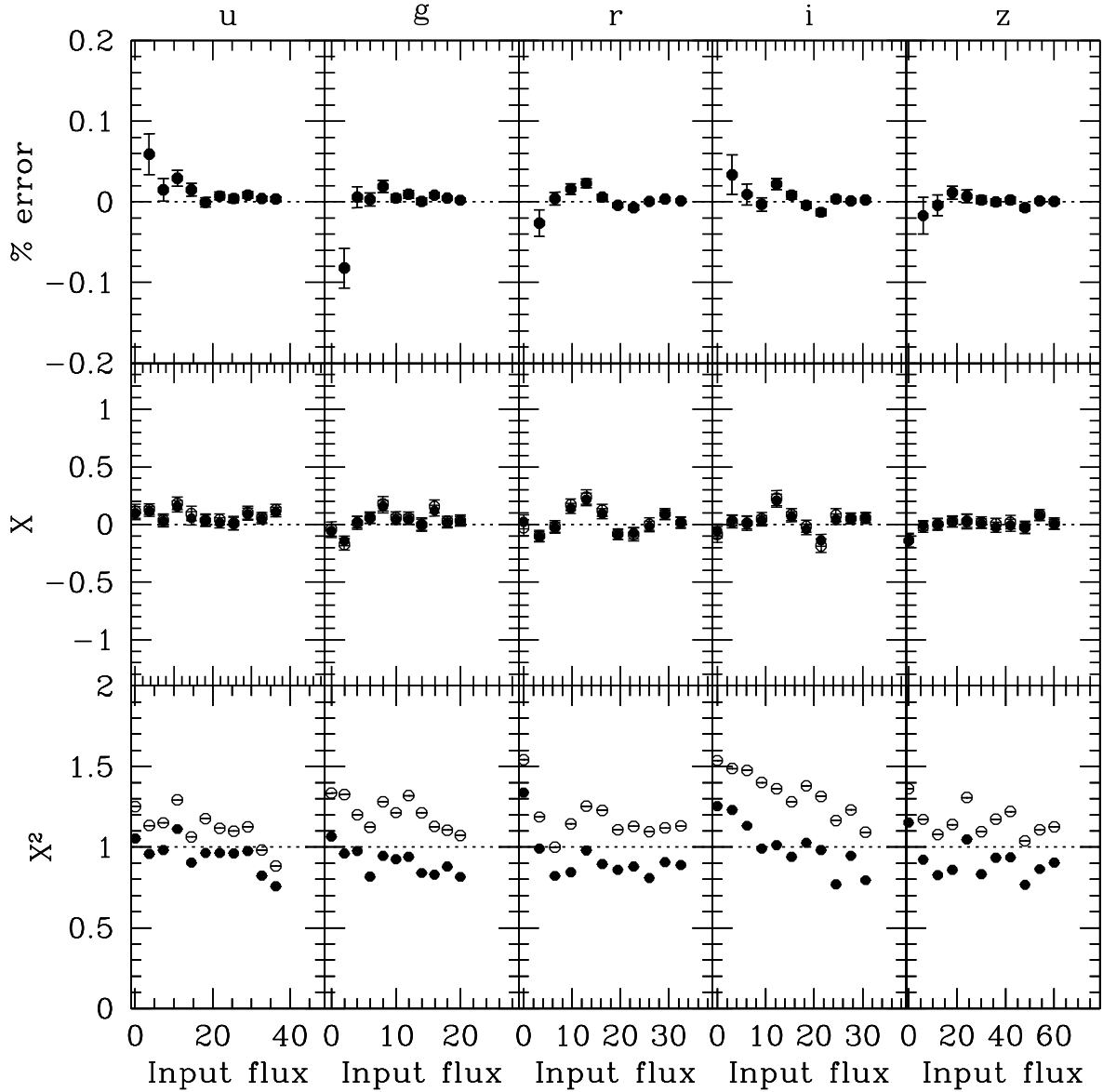}
\caption{Mean recovered fluxes of artificial supernovae at a range of
different input fluxes; the units
are $\mu Jy$ (note $m=20$ corresponds to 36.31 $\mu Jy$.  Each point
represents an average of several hundred artificial objects placed in
different 2005 epochs at the location of 2006 supernovae positions.
Top panels show the percent error of the recovered
flux. The middle panels show the error in the derived brightness normalized by
the predicted errors. The bottom panel shows the \chisq of the recovered
brightnesses; open points show the values computed using only the random
error derived for the point source brightnesses, while the solid points include
a term for errors in the sky level added in quadrature.}
\label{fig:fakesum} 
\end{figure}

For each artificial supernova test, we computed the median flux
offsets between the measured and the input value, and calculate the mean
fractional error of the recovered measurements. These are shown in the top
panel of Figure \ref{fig:fakesum} as a function of the input brightness.
The error bars are the computed error of the median values, given the
sample size.  One can see that the flux is recovered accurately: to within
a percent except for the faintest objects (and possibly even for these,
given the statistical errors). The few points that deviate the farthest from
a mean error of zero generally include locations where the artificial
star was placed at a location with a bright galaxy background (\ie in
the center of a galaxy). It is clear that this is the most challenging
situation for accurate recovery of a supernova brightness; if the
background is very bright, errors in the astrometry or PSF can throw
off the recovered supernova brightness.

Note that the error bars shown are the error of the sample mean; the
statistical error on individual measurements are much larger
than any small residual bias. This is demonstrated in the middle panel,
which shows the mean of the error in recovered magnitude, normalized by
the error estimate.  

The bottom panel of Figure \ref{fig:fakesum} shows a reduced $\chi^2$
value, calculated from the square of the difference between
recovered and input magnitude, normalized by the estimated variance.
If our error estimates are perfect, these should have values near
unity. The open points show $\chi^2$ as computed using the random error
on the derived intensities; in general, these are slightly larger than
unity. However, if one adds in quadrature the systematic errors from the
sky determination (which should be random over a set of observations
on different dates), then one gets the $\chi^2$ values shown with the
filled points. These show that using error estimate based on a combination
of the flux plus sky error gives accurate error estimates, although
our estimate of the sky error may be slightly too large.

\section{Light curves}

In the 2005 SDSS SN season, 130 type Ia supernovae were spectroscopically
confirmed, along with an additional 16 spectroscopically probable Type
Ia's.  
A complete list of all of the discovered supernovae, along with positions
and IAU designations, including non-Ia supernovae, is presented in 
Sako \etal (2008).


We have used scene modelling to derive light curves for the 146 objects;
the photometry data is available in the electronic version of this paper. 
Table \ref{tab:2005hk} 
shows a portion of a sample data table for one of our supernovae, SN2005hk,
which has been discussed by Phillips \etal (2007). 
The files contain
several lines of header information about the object: the SDSS internal
candidate ID number, the IAU designation, the position, SDSS type,
and redshift. In addition, approximate underlying galaxy surface
brightnesses in each bandpass are given, as determined by the scene
modelling photometry. The epoch of each observations is given as a
modified heliocentric Julian date.  The magnitudes in the file are given
as asinh magnitudes (Lupton \etal 1999) on the native SDSS photometric
system, using the softening parameters given in Stoughton \etal (2002).
The fluxes are given in units of microJanskys, using the corrections
to an AB system described in Section \ref{sect:fluxcal}; by definition,
an object with an AB magnitude of zero and flat $F_\nu$ spectrum has a
flux of $3.631 \times 10^9 \mu Jy$.  Although no extinction correction
has been applied to the measured brightnesses, the Galactic extinction as
estimated from the Schlegel, Finkbeiner, \& Davis (1998) maps is given in
the file headers. The spectroscopic observations and the determination of
the redshifts are described in Zheng \etal (2008).  The redshifts, which
were obtained by a variety of telescopes (Hobby-Eberly Telescope, Apache
Point Observatory 3.5-m, Subaru Telescope, William Herschel Telescope,
Nordic Optical Telescope, ESO New Technology Telescope, WIYN Telescope,
Keck Observatory, and the South African Large Telescope), are in the
heliocentric frame.

\begin{figure}
\plotone{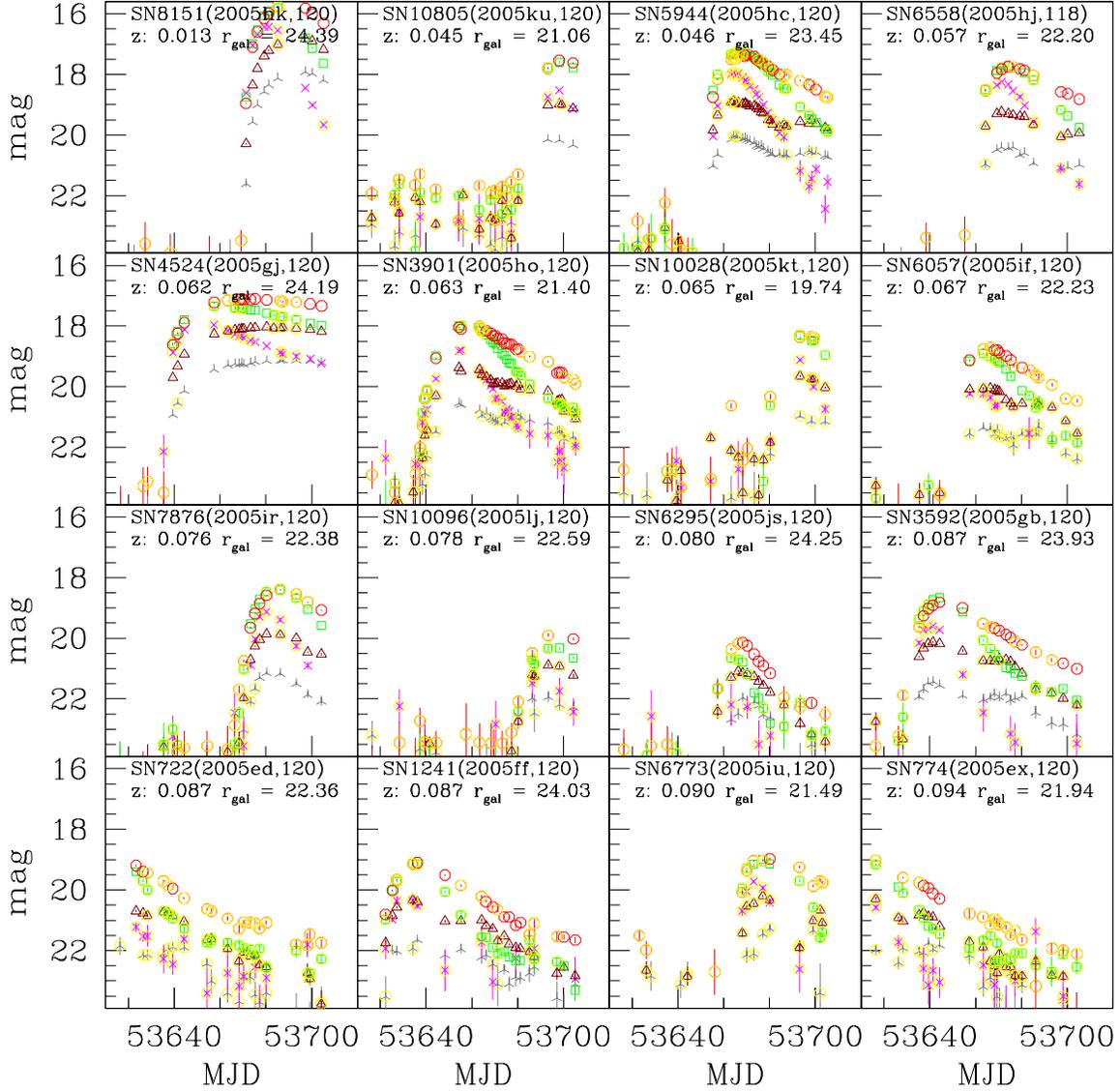}
\caption{Derived light curves for the 2005 type Ia supernovae, sorted
by redshift.  Red points are $r$, green points are $g$, magenta points
are $u$, brown points are $i+1$, and grey points are $z+2$. Points
circled in yellow have non-zero photometry flags; points with flag value 
greater than 1024 (see text) are not plotted. The SDSS type is given,
along with the IAU designation, in parentheses: type 120 are highly likely
type Ia SN confirmed by the SDSS survey team, type 119 objects are probably
type Ia's, and type 118 are Ia's confirmed by another team.}
\label{fig:lc1}
\end{figure}

In Figure \ref{fig:lc1},  we show our derived light curves for the 146
supernovae, sorted in order of redshift. These demonstrate the quality of
the light curves. The plots include information about the IAU designation
of these supernovae, and also give the estimated $r$-band galaxy surface
brightness (from the scene modelling results) at the location of the
supernova.


\section{Conclusion}

We have presented a general technique, scene-modelling photometry,
for extracting supernovae photometry from multiple observations. A key
feature of this technique is that it does not require resampling of data,
resulting in accurate photometry and error estimates. Fitting all of the
images as a sum of supernova and galaxy light results in optimal use
of all of the data, giving the highest precision in the determination
of the supernova light curves. Another important consequence of this
technique is that it is straightforward to combine data from several
pointings or even telescopes, although the existence of non-zero color
terms between different telescopes remains a limitation in the accuracy
of the photometry.

We use the technique to extract photometry for all of the confirmed and
probable type Ia supernova candidates from the 2005 SDSS SN season. All
of the data is accessible for public use via electronic tables and will
also be available through the SDSS supernova web site. These data provide
the basis for the initial analysis of the SDSS supernova survey.

Funding for the creation and distribution of the SDSS and SDSS-II has
been provided by the Alfred P. Sloan Foundation, the Participating
Institutions, the National Science Foundation, the U.S. Department of
Energy, the National Aeronautics and Space Administration, the Japanese
Monbukagakusho, the Max Planck Society, and the Higher Education Funding
Council for England. The SDSS Web site is http://www.sdss.org/.  
This work was also partially supported by Los Alamos National Laboratory 
University of California Directed Research and Development Fund.
The Hobby-Eberly Telescope (HET) is a joint project of the University of Texas
at Austin, the Pennsylvania State University,  Stanford University,
Ludwig-Maximillians-Universit\"at M\"unchen, and Georg-August-Universit\"at
G\"ottingen; the HET is named in honor of its principal benefactors,
William P. Hobby and Robert E. Eberly.

The SDSS is managed by the Astrophysical Research Consortium for the
Participating Institutions. The Participating Institutions are the
American Museum of Natural History, Astrophysical Institute Potsdam,
University of Basel, Cambridge University, Case Western Reserve
University, University of Chicago, Drexel University, Fermilab, the
Institute for Advanced Study, the Japan Participation Group, Johns
Hopkins University, the Joint Institute for Nuclear Astrophysics,
the Kavli Institute for Particle Astrophysics and Cosmology, the
Korean Scientist Group, the Chinese Academy of Sciences (LAMOST), Los
Alamos National Laboratory, the Max-Planck-Institute for Astronomy
(MPA), the Max-Planck-Institute for Astrophysics (MPiA), New Mexico
State University, Ohio State University, University of Pittsburgh,
University of Portsmouth, Princeton University, the United States Naval
Observatory, and the University of Washington.

\newpage

\begin{deluxetable}{llllll}
\tablecaption{Observed and synthetic measurements of solar analogs}
\tablewidth{0pt}
\tablehead{
\colhead{star}&\colhead{\textit{u}}&\colhead{\textit{g}}&\colhead{\textit{r}}&\colhead{\textit{i}}&\colhead{\textit{z}}
}
\startdata
\multicolumn{6}{l}{Observed mags (transformed to SDSS)}\\
P330E & 14.548 & 13.280 & 12.841 & 12.701 & 12.674 \\
P177D & 15.118 & 13.745 & 13.300 & 13.158 & 13.125 \\
P041C & 13.573 & 12.260 & 11.844 & 11.719 & 11.703 \\
Synthetic mags&&&&&\\
P330E & 14.506 & 13.303 & 12.839 & 12.708 & 12.675 \\
P177D & 15.085 & 13.776 & 13.307 & 13.178 & 13.142 \\
P041C & 13.537 & 12.279 & 11.852 & 11.746 & 11.732 \\
Differences ($m_{syn} - m_{obs}$)&&&&&\\
$\Delta m$& -0.037 & 0.024 & 0.005 & 0.018 & 0.016 \\
rms $\Delta m$ & 0.005 & 0.006 & 0.005 & 0.010 & 0.014 \\
\enddata
\label{tab:ab}
\end{deluxetable}

\begin{deluxetable}{lllllllllllllll}
\rotate
\tabletypesize{\scriptsize}
\tablewidth{8.75in}
\tablecaption{Photometry for SN 2005hk (SDSS SN 8151)}
\tablehead{
\colhead{FLAG\tablenotemark{a}}&\colhead{MJD}&\colhead{FILT\tablenotemark{b}}&\colhead{MAG\tablenotemark{c}}&\colhead{MERR\tablenotemark{d}}&\colhead{MSERR\tablenotemark{e}}&\colhead{MGERR\tablenotemark{f}}&\colhead{FLUX\tablenotemark{g}}&\colhead{FLUXERR\tablenotemark{h}}&\colhead{SERR\tablenotemark{i}}&\colhead{GERR\tablenotemark{j}}&\colhead{NPRE\tablenotemark{k}}&\colhead{TELE}&\colhead{RUN\tablenotemark{l}}&\colhead{STRIP\tablenotemark{m}}}
\startdata
0&53671.34315&1&18.745&0.012&0.001&0.001&1.128E+02&1.247E+00&1.479E-01&9.707E-02&10&sdss&5786&82S\\
0&53671.33983&2&18.960&0.018&0.005&0.001&9.419E+01&1.562E+00&4.341E-01&1.051E-01&12&sdss&5786&82S\\
0&53671.34066&3&19.288&0.023&0.003&0.002&6.880E+01&1.458E+00&2.055E-01&1.349E-01&12&sdss&5786&82S\\
0&53671.34232&4&19.609&0.096&0.066&0.012&5.115E+01&4.547E+00&3.125E+00&5.661E-01&12&sdss&5786&82S\\
0&53671.34149&0&18.612&0.035&0.003&0.003&1.349E+02&4.349E+00&3.877E-01&3.688E-01&10&sdss&5786&82S\\
0&53674.24276&1&16.989&0.012&0.000&0.000&5.686E+02&6.285E+00&1.035E-01&9.707E-02&10&sdss&5797&82S\\
0&53674.23944&2&17.103&0.006&0.000&0.000&5.210E+02&2.879E+00&1.724E-01&1.051E-01&12&sdss&5797&82S\\
0&53674.24027&3&17.352&0.009&0.001&0.000&4.093E+02&3.393E+00&4.193E-01&1.349E-01&12&sdss&5797&82S\\
0&53674.24193&4&17.576&0.017&0.004&0.002&3.336E+02&5.224E+00&1.132E+00&5.661E-01&12&sdss&5797&82S\\
0&53674.24110&0&17.044&0.023&0.001&0.001&5.718E+02&1.211E+01&2.840E-01&3.688E-01&10&sdss&5797&82S\\
0&53676.33207&1&16.523&0.004&0.000&0.000&8.734E+02&3.218E+00&7.615E-02&9.707E-02&10&sdss&5807&82S\\
0&53676.32875&2&16.598&0.004&0.000&0.000&8.295E+02&3.056E+00&7.532E-02&1.051E-01&12&sdss&5807&82S\\
0&53676.32958&3&16.811&0.005&0.001&0.000&6.736E+02&3.102E+00&3.147E-01&1.349E-01&12&sdss&5807&82S\\
0&53676.33124&4&17.016&0.010&0.003&0.001&5.587E+02&5.146E+00&1.419E+00&5.661E-01&12&sdss&5807&82S\\
0&53676.33041&0&16.675&0.014&0.001&0.000&8.032E+02&1.036E+01&3.986E-01&3.688E-01&10&sdss&5807&82S\\
\multicolumn{15}{l}{...}
\enddata
\tablecomments{Data files for each supernova are published in their entirety in the electronic edition of the {\it Astronomical Journal}.  A portion of a table is shown here for guidance regarding its form and content. The online files include some additional ancillary information about each object, including the IAU designation, the coordinates, the redshift, the expected foreground extinctions from Schlegel et al., and the derived underlying galaxy brightnesses from the scene modelling.}
\tablenotemark{a}{For details of (bitwise) values see Holtzman et al. (2008).  A value of 0 indicates no lines, $>1024$ is very likely a bad measurement, while a value between 0 and 1024 is likely OK but frame not used for galaxy solution.}
\tablenotemark{b}{01234 = ugriz bands.}
\tablenotemark{c}{MAG is in native SDSS photometric system, and is an asinh magnitude. No extinction correction has been applied.}
\tablenotemark{d}{Random error in magnitude.}
\tablenotemark{e}{Systematic magnitude error estimate from error in sky estimate.}
\tablenotemark{f}{Systematic magnitude error estimate from error in underlying galaxy brightness.}
\tablenotemark{g}{FLUX is in microJy using SDSS/AB correction from Holtzman et al. (2008).}
\tablenotemark{h}{Random error in flux.}
\tablenotemark{i}{Systematic flux error estimate from error in sky estimate.}
\tablenotemark{j}{Systematic flux error estimate from error in underlying galaxy brightness.}
\tablenotemark{k}{RUN gives the SDSS run identifier.}
\tablenotemark{l}{Strip gives the SDSS strip for this measurement.}
\tablenotemark{m}{NPRE gives the number of pre-SN observations used.}
\end{deluxetable}

%
%
%
%

\newpage

\clearpage

\begin{figure}
\plotone{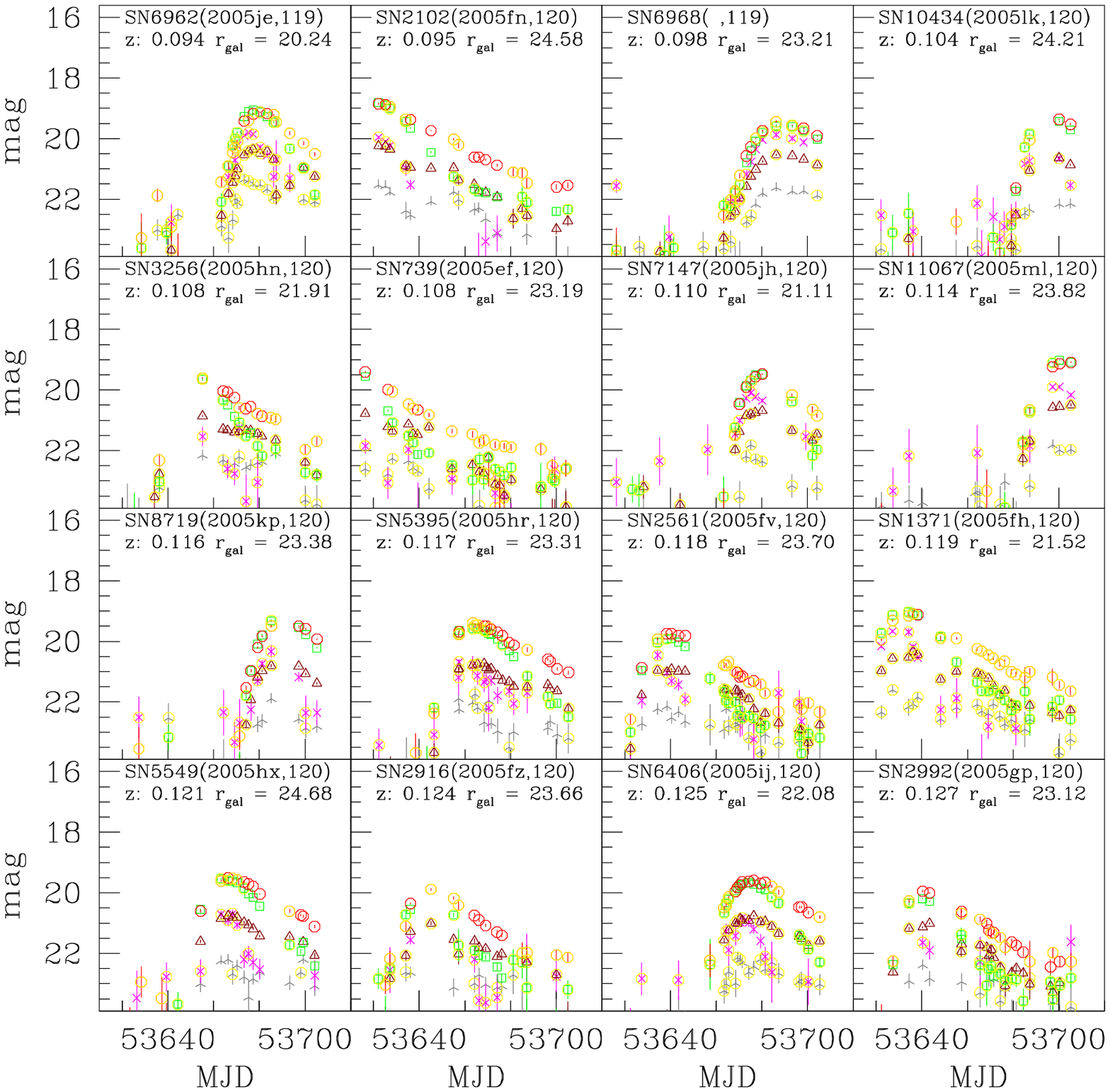}
\label{fig:lc2}
\end{figure}

\clearpage
\begin{figure}
\plotone{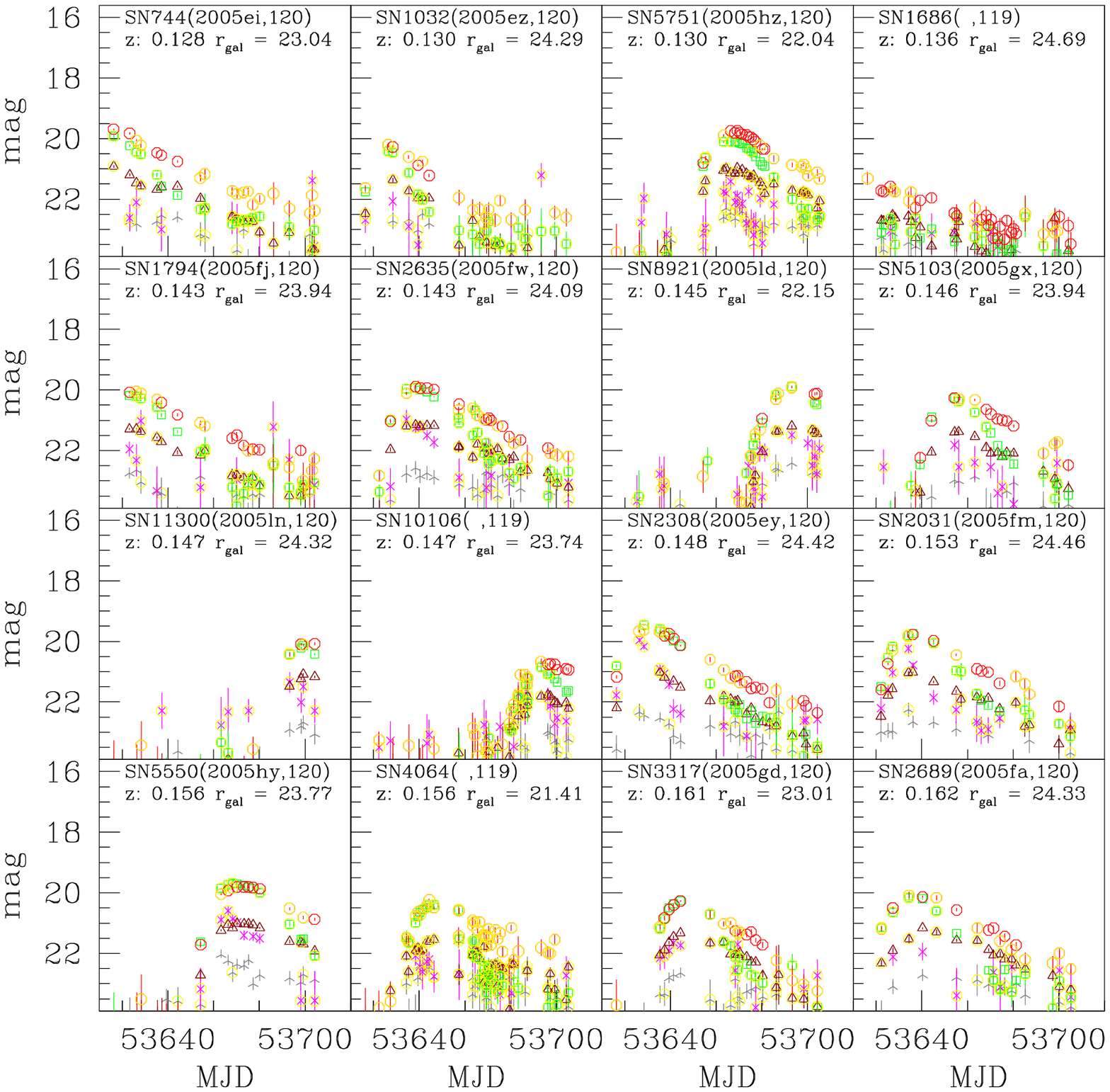}
\label{fig:lc3}
\end{figure}

\clearpage
\begin{figure}
\plotone{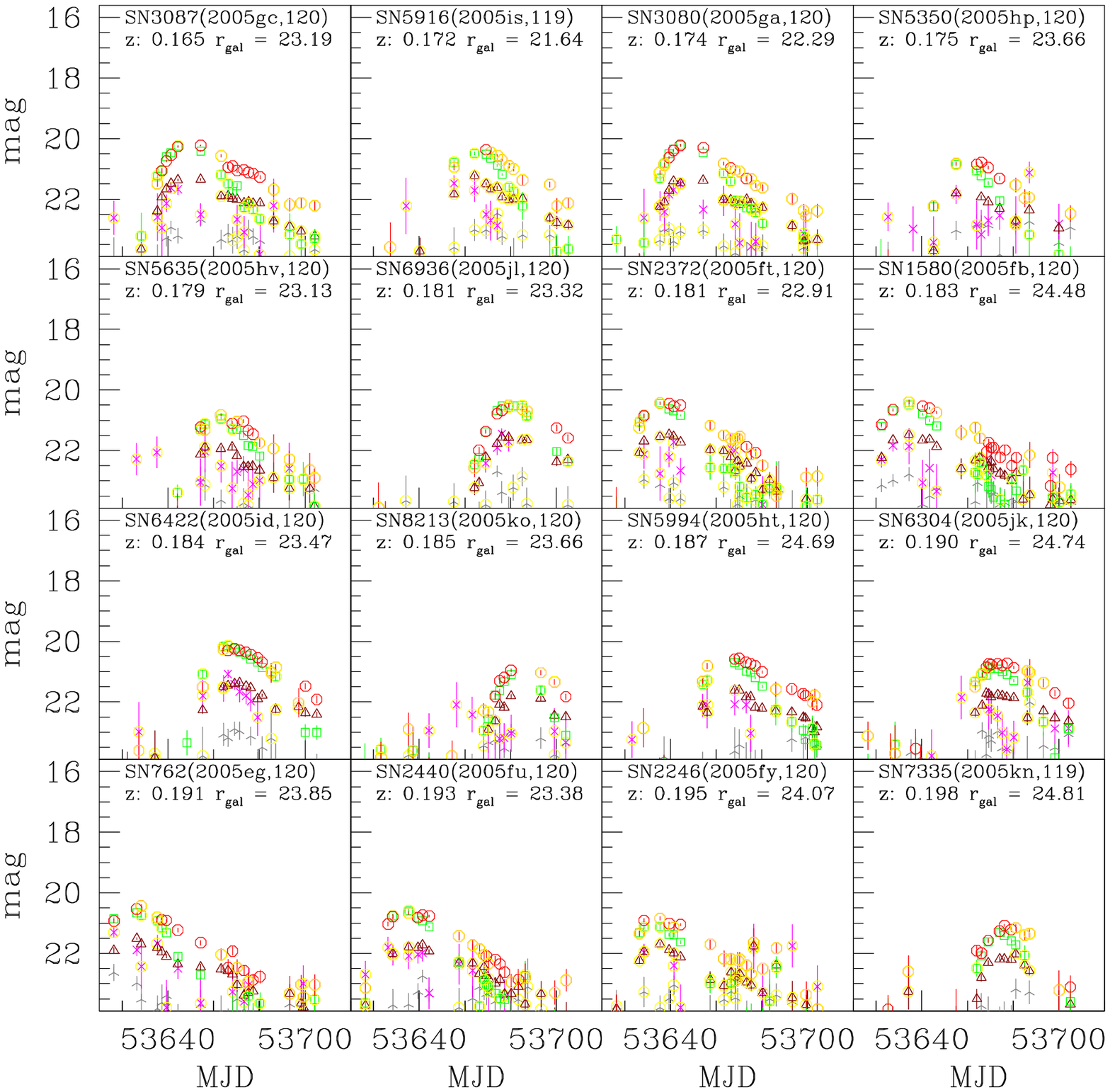}
\label{fig:lc4}
\end{figure}

\clearpage
\begin{figure}
\plotone{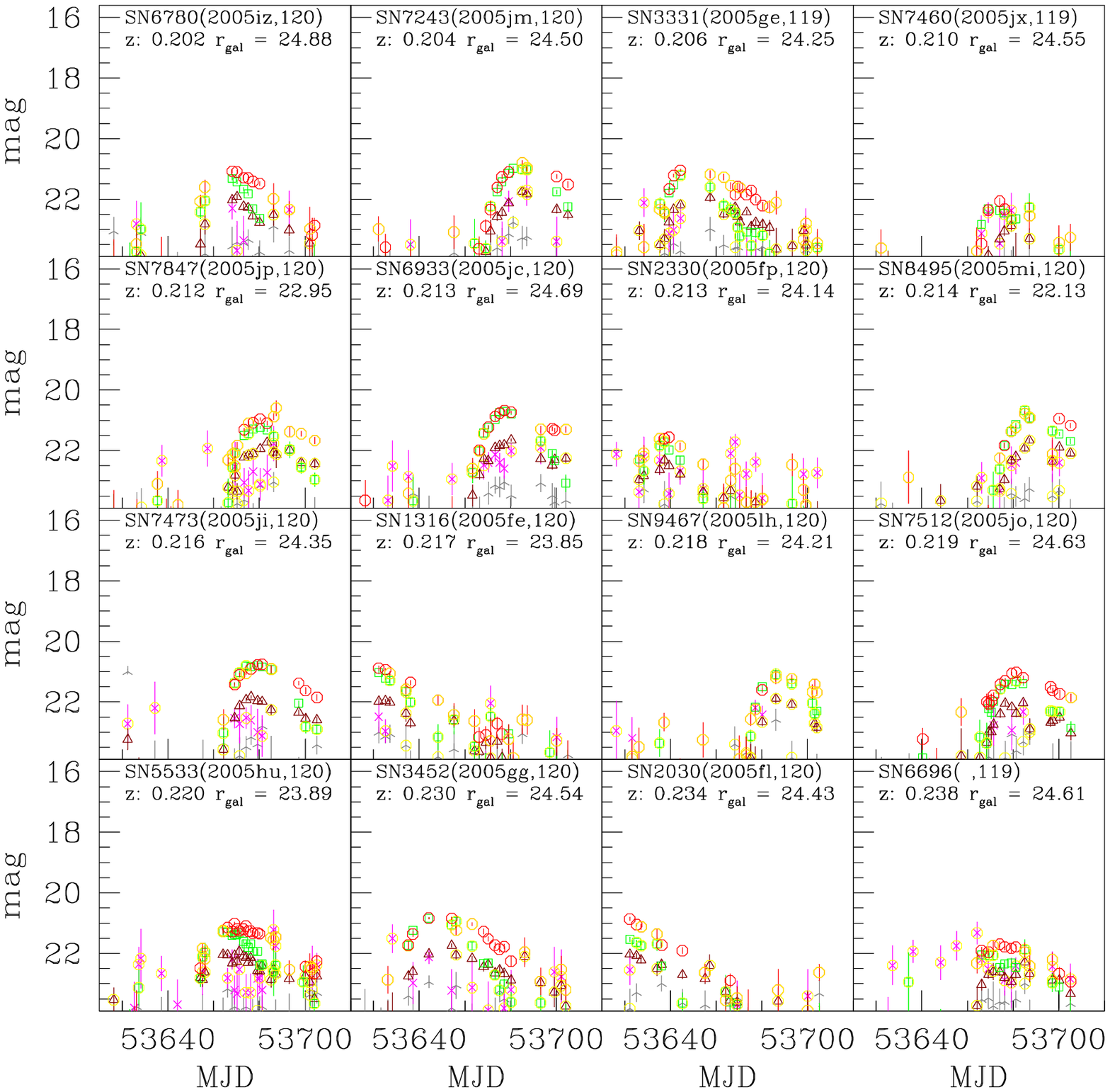}
\label{fig:lc5}
\end{figure}

\clearpage
\begin{figure}
\plotone{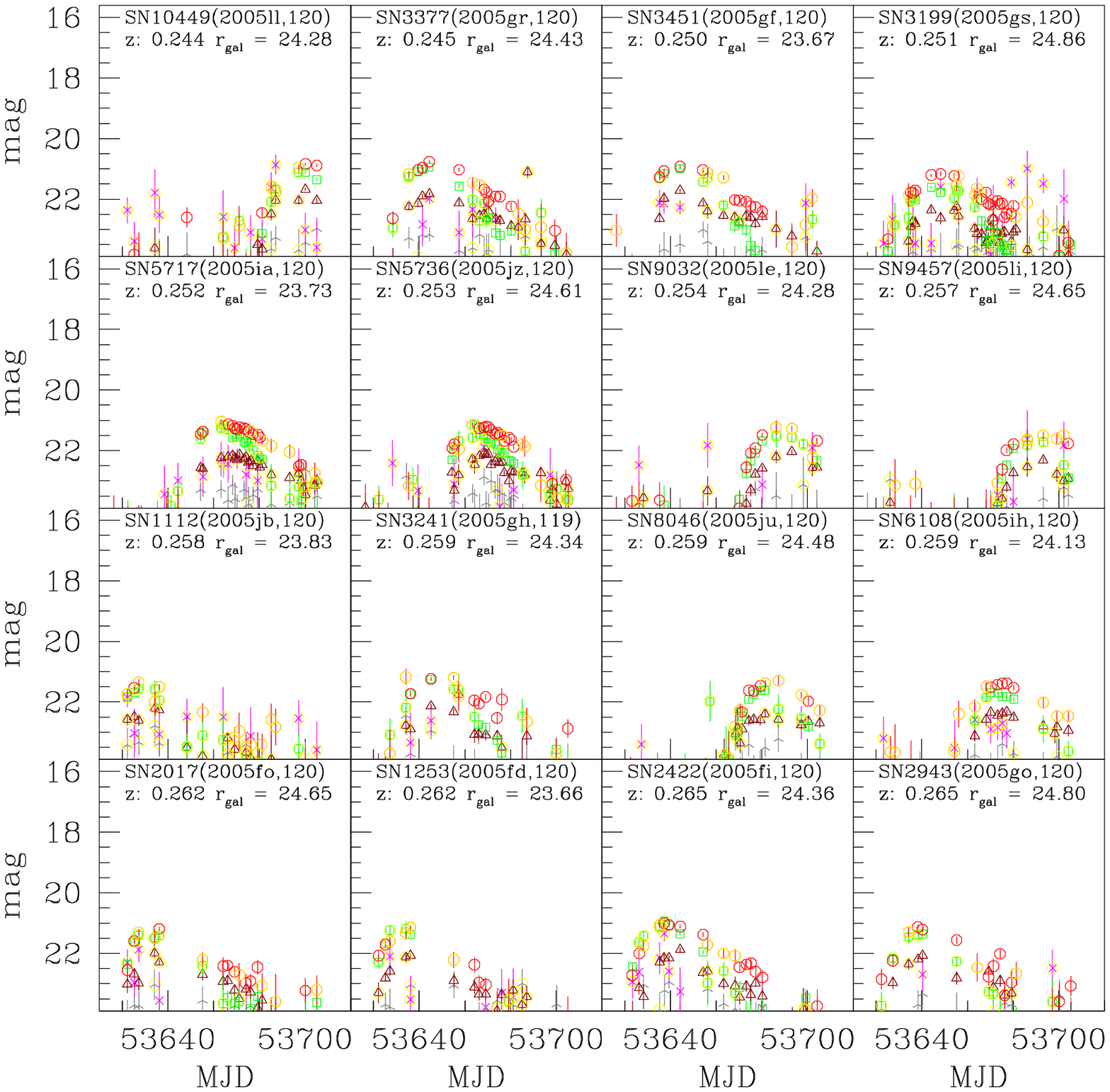}
\label{fig:lc6}
\end{figure}

\clearpage
\begin{figure}
\plotone{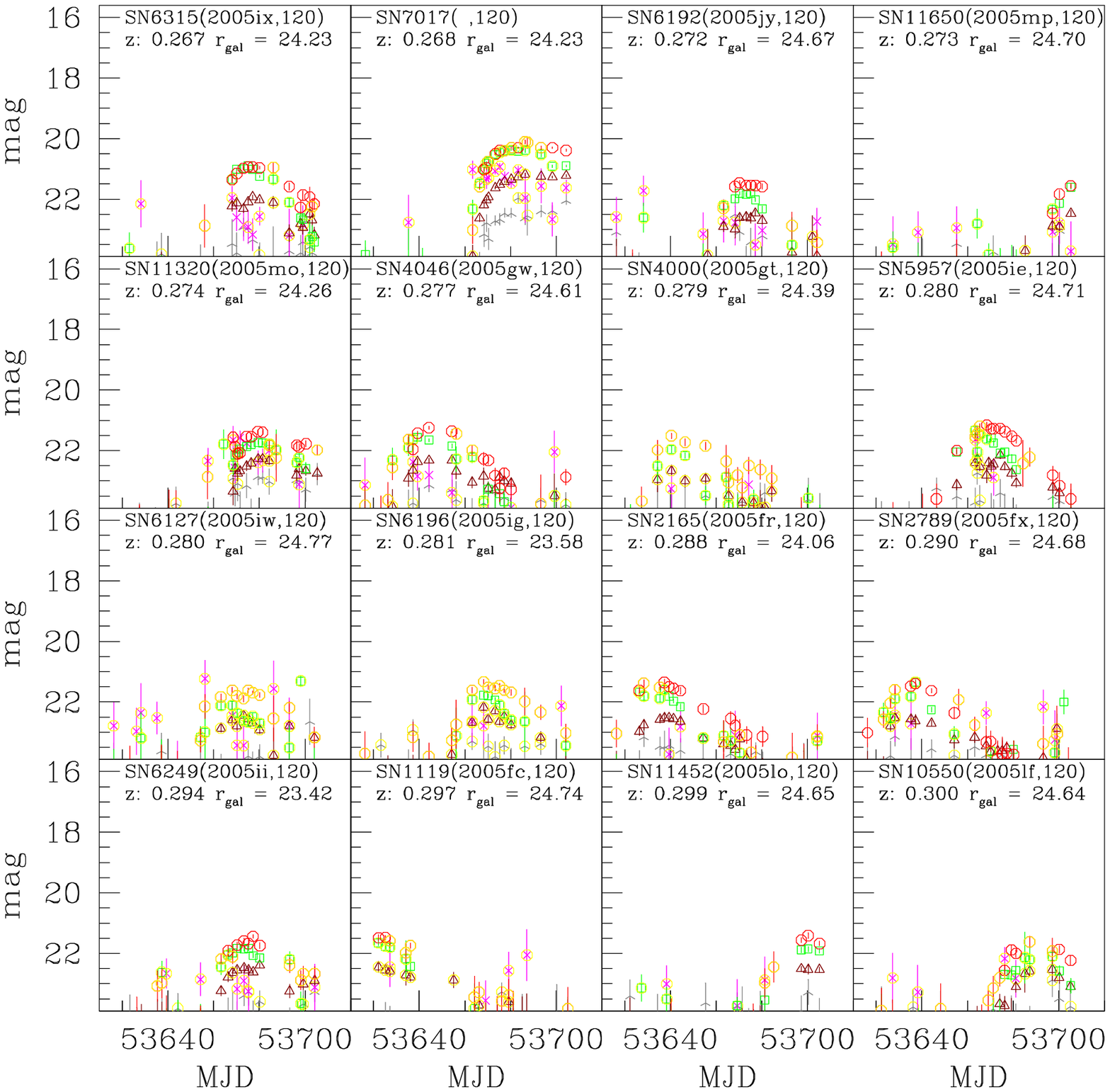}
\label{fig:lc7}
\end{figure}

\clearpage
\begin{figure}
\plotone{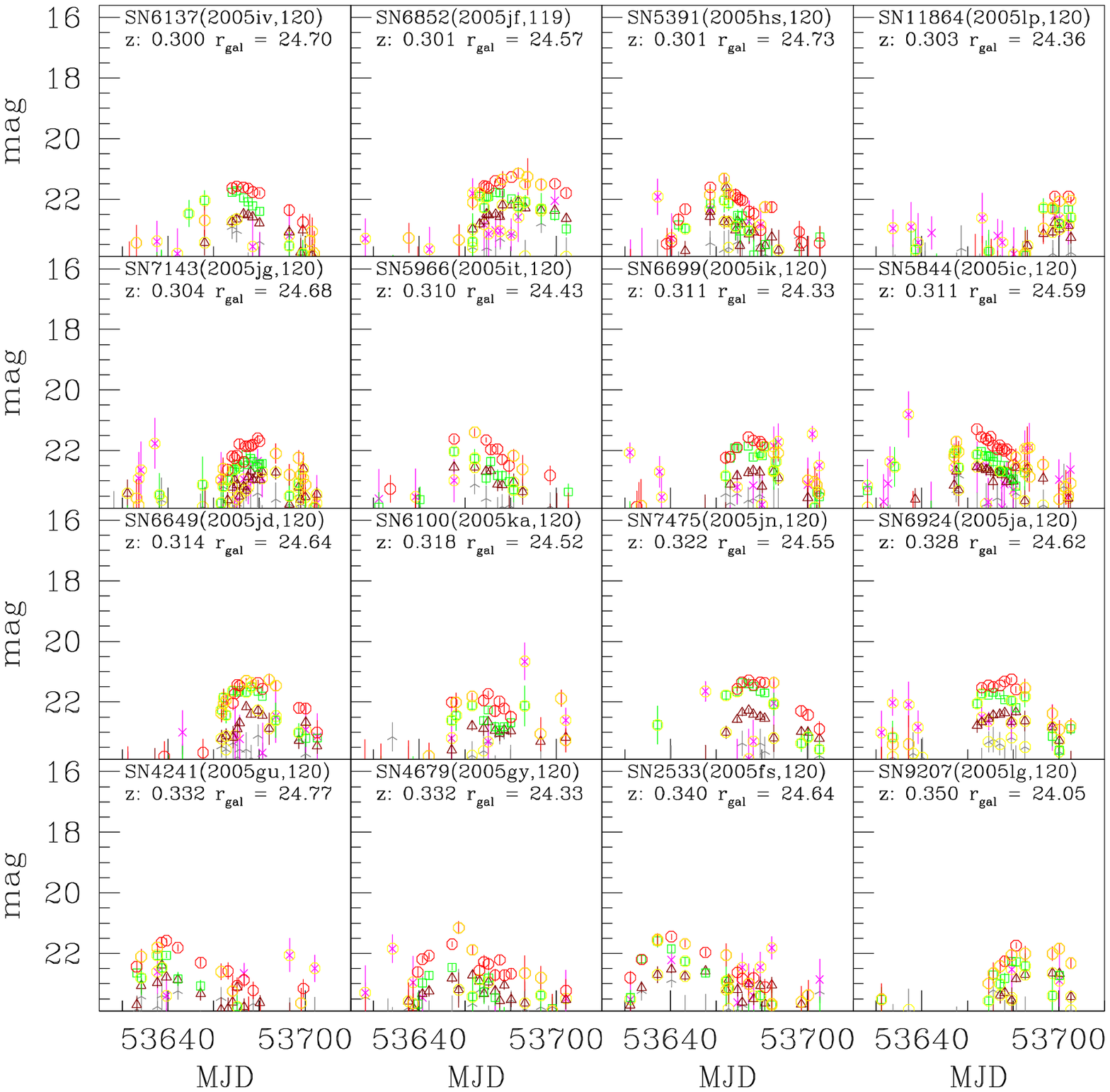}
\label{fig:lc8}
\end{figure}

\clearpage
\begin{figure}
\plotone{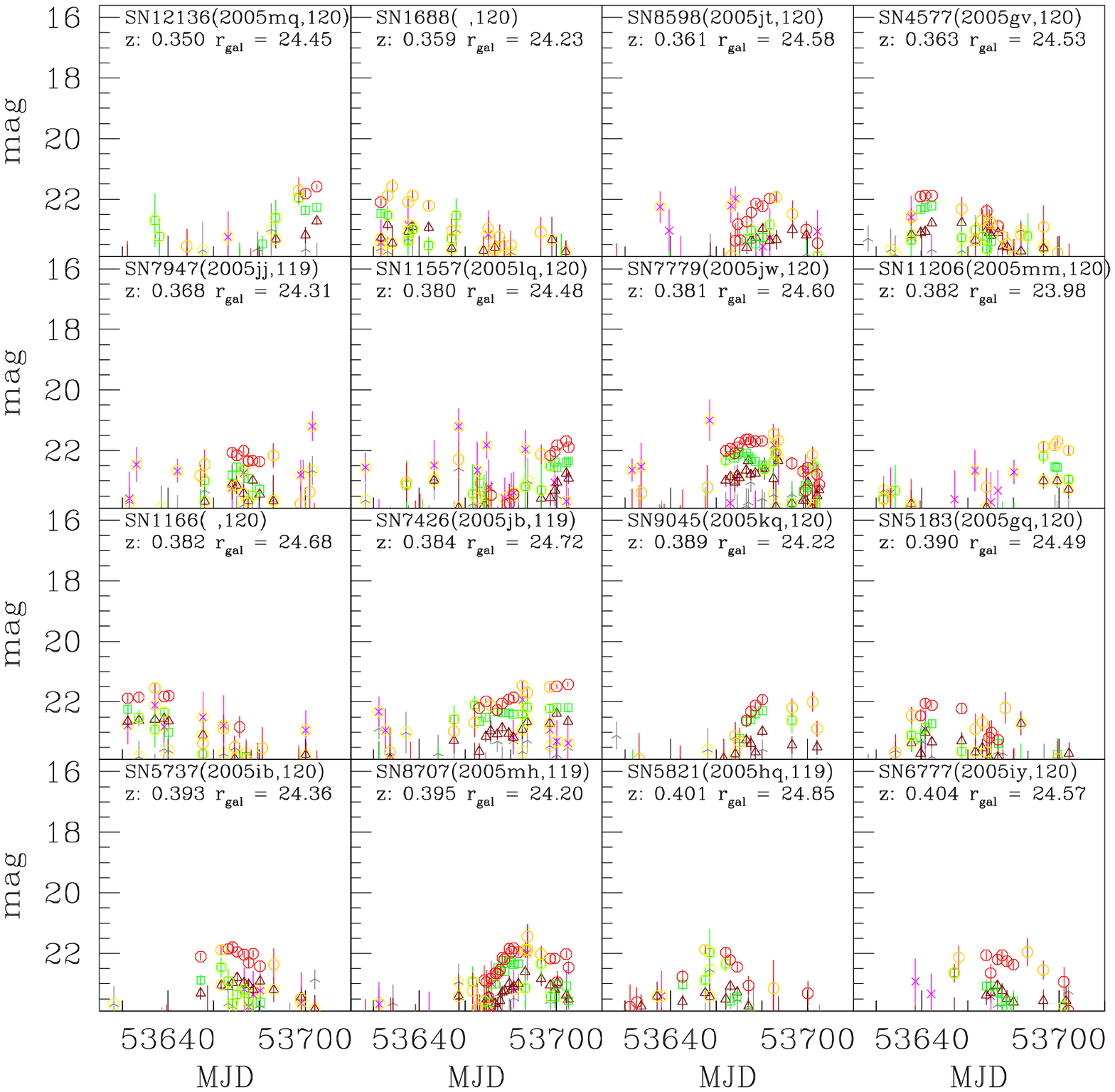}
\label{fig:lc9}
\end{figure}

\clearpage
\begin{figure}
\plotone{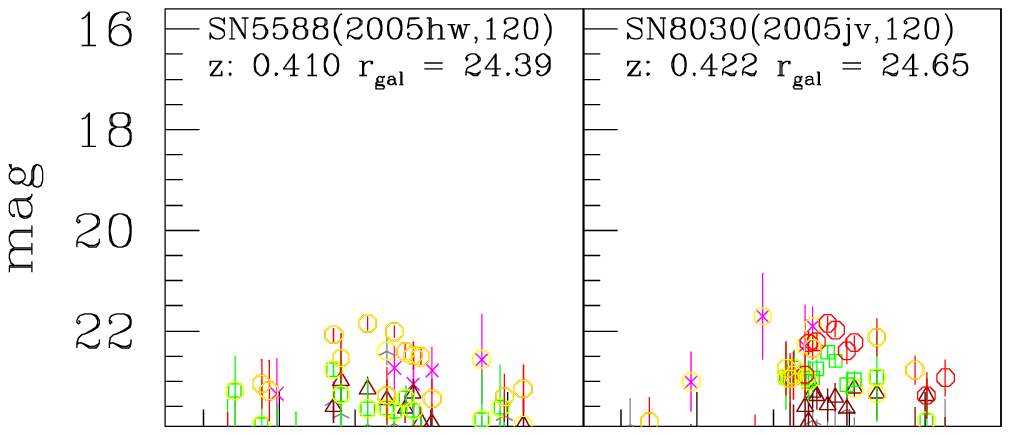}
\label{fig:lc10}
\end{figure}

\end{document}